%% file: main.tex
\documentclass[a4paper]{cas-dc}

\usepackage[numbers]{natbib}
\usepackage{hyperref}
\usepackage{float}
\usepackage{amsmath, nccmath}  
\usepackage{caption}
\usepackage{lipsum}
\usepackage{blindtext}
\usepackage[mathscr]{eucal}
\usepackage{url}

\begin{document}
\captionsetup{justification = centering, singlelinecheck = false}

\let\WriteBookmarks\relax
\def\floatpagepagefraction{1}
\def\textpagefraction{.001}
\shorttitle{HATS: A Hierarchical Graph Attention Network for Stock Movement Prediction}
\shortauthors{Raehyun Kim et~al.}

\title [mode = title]{HATS: A Hierarchical Graph Attention Network for Stock Movement Prediction}       

\author[add1]{Raehyun Kim}
\author[add2]{Chan Ho So}
\author[add1]{Minbyul Jeong}
\author[add1]{Sanghoon Lee}
\author[add1]{Jinkyu Kim}
\author[add1, add2]{Jaewoo Kang}
\cormark[1]

\address[add1]{Department of Computer Science and Engineering, Korea University, 145, Anam-ro, Seongbuk-gu, Seoul, South Korea, 02841}
\address[add2]{Interdisciplinary Graduate Program in Bioinformatics, Korea University, 145, Anam-ro, Seongbuk-gu, Seoul, South Korea, 02841}

\cortext[cor1]{Corresponding author\\ 
 E-mail address: kangj@korea.ac.kr (J. Kang), raehyun@korea.ac.kr (R. Kim), chanhoso@korea.ac.kr (C. So), minbyuljeong@korea.ac.kr(M. Jeong), a11525@korea.ac.kr(S. Lee), no100kill@korea.ac.kr(J. Kim)}


\begin{abstract}
Many researchers both in academia and industry have long been interested in the stock market. Numerous approaches were developed to accurately predict future trends in stock prices. Recently, there has been a growing interest in utilizing graph-structured data in computer science research communities. Methods that use relational data for stock market prediction have been recently proposed, but they are still in their infancy. First, the quality of collected information from different types of relations can vary considerably. No existing work has focused on the effect of using different types of relations on stock market prediction or finding an effective way to selectively aggregate information on different relation types. Furthermore, existing works have focused on only individual stock prediction which is similar to the node classification task. 

To address this, we propose a hierarchical attention network for stock prediction (HATS) which uses relational data for stock market prediction. Our HATS method selectively aggregates information on different relation types and adds the information to the representations of each company. Specifically, node representations are initialized with features extracted from a feature extraction module. HATS is used as a relational modeling module with initialized node representations. Then, node representations with the added information are fed into a task-specific layer. Our method is used for predicting not only individual stock prices but also market index movements, which is similar to the graph classification task. The experimental results show that performance can change depending on the relational data used. HATS which can automatically select information outperformed all the existing methods. 

\end{abstract}


\maketitle

\input{sections/Introduction.tex}

\input{sections/Preliminary.tex}

\input{sections/Methodology.tex}
\input{sections/Data.tex}

\input{sections/Experiments.tex}

\input{sections/Conclusion.tex}

\section*{Acknowledgement}
This work was supported by the National Research Foundation of
Korea (NRF-2017R1A2A1A17069645, 

NRF-2017M3C4A7065887)


\bibliographystyle{cas-model2-names}

\bibliography{Ref}

\newpage
\bio{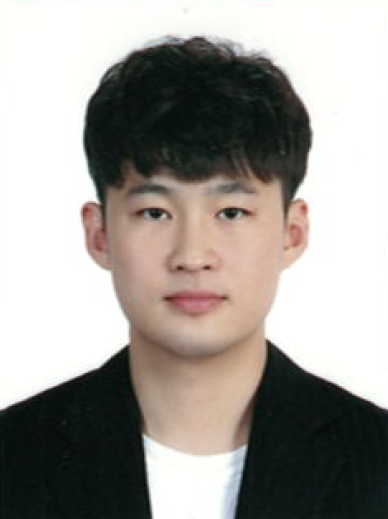}
\textbf{Raehyun Kim}
received a B.S. degree in business from Korea University, Seoul, Republic of Korea, in 2018, where he is currently pursuing an M.S. degree in computer science. His current research interests include financial market prediction and recommendation systems.
\vspace{4\baselineskip}
\endbio

\bio{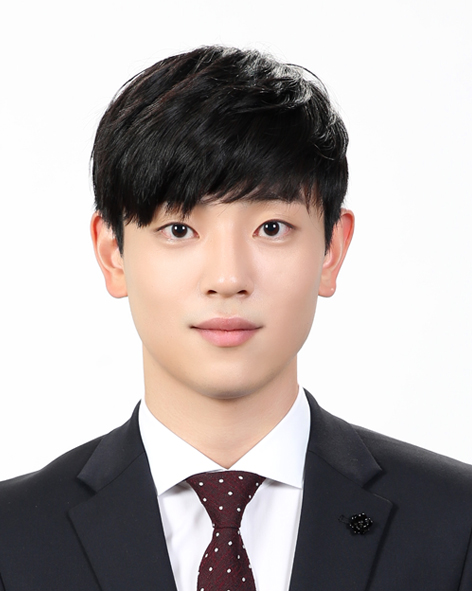}
\textbf{Chan Ho So}
received a B.S. degree in computer science from Korea University, Seoul, Republic of Korea, in 2018, where he is currently pursuing an M.S. degree in computer science. His current research interests include bioinformatics and named entity recognition.
\vspace{4\baselineskip}
\endbio

\bio{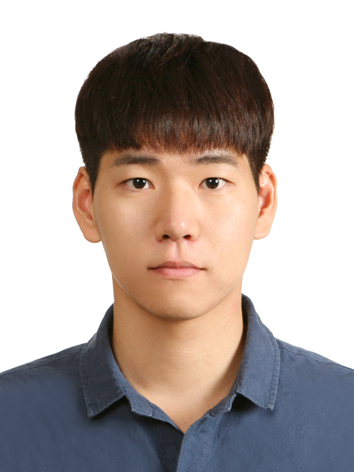}
\textbf{Minbyul Jeong}
received a B.S. degree in computer science from Korea University, Seoul, Republic of Korea, in 2019, where he is currently pursuing an M.S. degree in computer science. His current research interests include natural language processing and bioinformatics.
\vspace{4\baselineskip}
\endbio

\bio{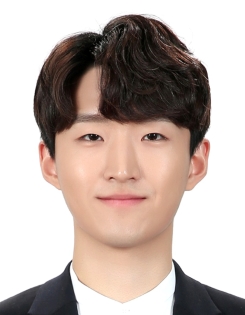}
\textbf{Sanghoon Lee}
received a B.S. degree in computer science from Korea University, Seoul, Republic of Korea, in 2019, where he is currently pursuing an M.S. degree in computer science. His current research interests include  bioinformatics.
\vspace{4\baselineskip}
\endbio

\bio{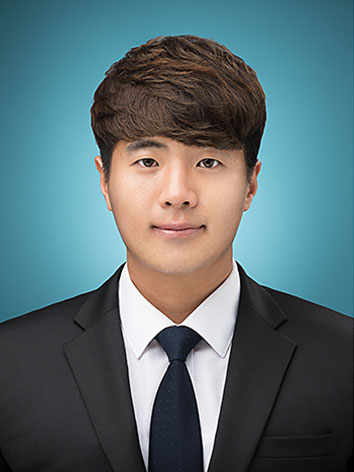}
\textbf{Jinkyu Kim}
received a B.S. degree in business from Korea University, Seoul, Republic of Korea, in 2019. His current research interests include financial market prediction and bioinformatics.
\vspace{4\baselineskip}
\endbio

\bio{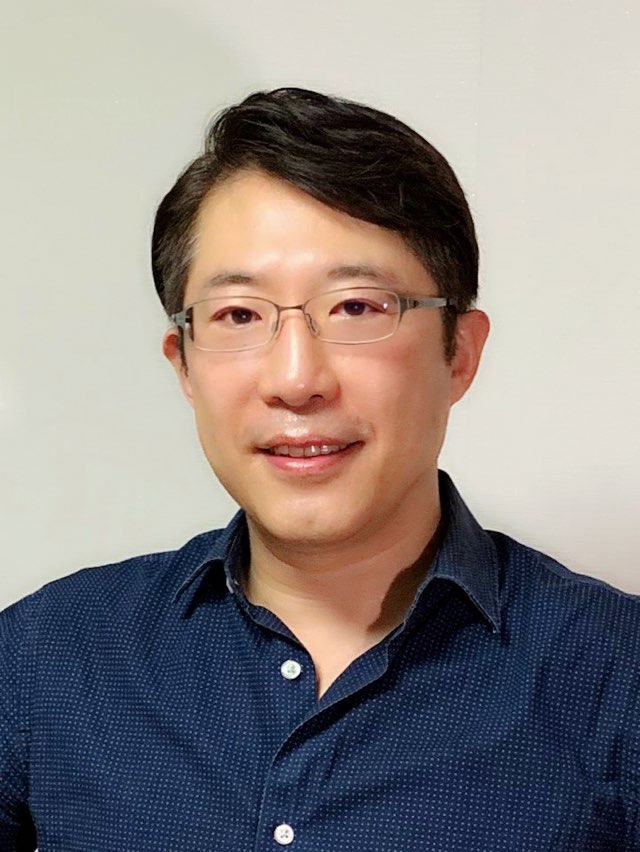}
\textbf{Jaewoo Kang} received a B.S. degree in computer science from Korea University, Seoul, Republic of Korea, in 1994, an M.S. degree in computer science from the University of Colorado Boulder, CO, USA, in 1996, and a Ph.D. degree in computer science from the University of Wisconsin-Madison, WI, USA, in 2003.

From 1996 to 1997, he was a Technical Staff Member at AT\&T Labs Research, Florham Park, NJ, USA. From 1997 to 1998, he was a Technical Staff Member at Savera Systems Inc., Murray Hill, NJ, USA. From 2000 to 2001, he was the CTO and Founder of WISEngine Inc., Santa Clara, CA, USA, and Seoul, Republic of Korea. From 2003 to 2006, he was an Assistant Professor in the Department of Computer Science, North Carolina State University, Raleigh, NC, USA. Since 2006, he has been a Professor in the Department of Computer Science, Korea University, Seoul, Republic of Korea. He also serves as the Department Head for the Interdisciplinary Graduate Program in Bioinformatics at Korea University, Seoul, Republic of Korea. He is jointly appointed with the Department of Medicine, School of Medicine, Korea University, Seoul, Republic of Korea.
\vspace{2\baselineskip}

\endbio

\newpage
\appendix
\input{sections/Appendix.tex}

\end{document}

%% file: sections/Introduction.tex
\section{Introduction\label{sec:intro}}
\bigbreak
Stock markets are a symbol of market capitalism and billions of shares of stock are traded every day. In 2018, stocks worth more than 65 trillion U.S. dollars were traded worldwide and market capitalization of domestic companies listed in the U.S. exceeds the country's GDP \footnote{https://data.worldbank.org/}. Although stock movement prediction is a difficult problem, its solutions can be applied to industry. Many researchers in both industry and academia have long shown interest in predicting future trends in the stock market. Researchers focused on finding profitable patterns in historical data are known as quants in the financial industry and referred to as data scientists in general. Regardless of which term is used, such researchers are increasingly using more systematic trading algorithms to automatically make trading decisions.

Even though there is still room for debate \cite{EMH}, numerous studies have showed that the stock market is predictable to some extent \cite{bollerslev2014stock}, \cite{phan2015stock}. Existing methods are based on the ideas of fundamentalists or technicians, both of whom have different perspectives on the market. 

Fundamentalists believe that the price of securities of a company corresponds to the intrinsic value of the company or entity \cite{dechow2001short}. If the current price of a company's stock is lower than its intrinsic value, investors should buy the stock as its price will go up and eventually be the same as its fundamental value. The fundamental analysis of a company involves an in-depth analysis of its performance and profitability. The intrinsic value of the company is based on its product sales, employees, infrastructure, and profitability of its investments \cite{agrawal2013state}. 

Technicians, on the other hand, do not consider real world events when predicting future trends in the stock market. For technicians, stock prices are considered as only typical time series data with complex patterns. With appropriate preprocessing and modeling, patterns can be analyzed, from which profitable patterns may be extracted. The information used for technical analysis consists of mainly closing prices, returns, and volumes. The movement of stock prices is known to be stochastic and non-linear. Technical analysis studies focus on reducing stochasticity and capturing consistent patterns.

Some technical analysis works have focused on how to extract meaningful features from raw price data. In the finance industry, features extracted from such data are called technical indicators and include adaptive moving average, relative strength index, stochastics, momentum oscillator, and commodity channel index. Creating meaningful technical indicators is similar to manual feature engineering in general machine learning tasks. Like in any machine learning task, extracted features contain important information for models. Hence, some works have utilized indicators to more accurately predict the movement of stock prices \cite{dempster2001computational}, \cite{patel2015predicting}.

Technicians are also interested in finding meaningful patterns in raw price data. Numerous works have analyzed the effectiveness of different models. Although the majority of researchers agree that the stock market moves in a non-linear way, many empirical studies show that non-linear models do not outperform linear models \cite{adebiyi2014comparison}, \cite{agrawal2013state}. The results of these studies show that even though deep neural network based models have been successfully applied to many challenging domains, careful consideration should be given when trying to design profitable models for stock market prediction. Lately, more studies are finding that non-linear models outperform linear models \cite{rather2015recurrent}. Many studies have shown that recurrent neural network based models are effective in stock movement prediction \cite{bao2017deep}. 

As the amount of information on the web continues to rapidly increase, it becomes easier to obtain information on securities from different sources. Data scientists with a computer science background have begun to pay attention to unstructured data such as text from the news and Twitter \cite{bollen2011twitter}, \cite{ding2015deep}. Models that use text data reflecting the real world events of companies can be categorized as fundamental analysis models. However, text-based stock prediction approaches try to capture investors' opinions about an event. Based on the assumption that the price of a company's stock can be based on the total aggregation of investors' opinions about the company, some works focused on reading investor's opinions about companies \cite{li2014news}. There also exist researches focusing on understanding the impact of events on stock price \cite{ding2016knowledge}.

More recently, computer science research communities have been highly interested in utilizing graph-structured data \cite{GraphSAGE}, \cite{DiffPool}. Stock market prediction methods using corporate relational data have also been proposed \cite{CIKM-GraphStock}, \cite{Temporal}.  \citeauthor{CIKM-GraphStock} created a network of companies based on financial investment information \cite{CIKM-GraphStock}. Using a constructed adjacency matrix, they trained a GCN model and compared its prediction performance with that of more conventional network embedding models'. \citeauthor{Temporal} developed a more general framework \cite{Temporal} that involves using many different types of relations in a publicly available knowledge database. They also proposed a GNN model that can capture temporal features of stocks. Although these models were the first to integrate relational data for stock market prediction, they can be still be improved. The quality of information varies considerably depending on the type of relation. However, no existing work has thoroughly investigated which types of relational data are more beneficial to stock movement prediction or focus on finding an effective way to selectively aggregate information on different relation types. 

Furthermore, previous works have focused mainly on node classification. Node classification and graph classification are the two main tasks in graph-based learning. In a stock market network, individual nodes typically represent companies. Predicting future trends in individual stock prices is similar to the node classification task. We argue that previously proposed models can be used as a node representation updating function in the graph classification task which we propose in this work. 

To address the limitations mentioned above, in this paper, we study how to effectively utilize graph-based learning methods and relational data in stock market prediction. We use different types of relations and investigated their effect on performance in stock price movement prediction of individual companies. In our experiments, we found that only relevant relations are useful for stock prediction. Information from some irrelevant relations even degraded prediction performance. We propose HATS which is a new hierarchical graph attention network method that uses relational data for stock market prediction. HATS selectively aggregates information from different relations and adds the information to the representations of companies. Specifically, node features are initialized with extracted features from the feature extraction module. HATS is used as relational modeling module to selectively gather information from neighboring nodes. The representations with the added information are then fed into a task-specific prediction layer. 

We applied our method to the following two graph related tasks: predicting the movement of individual stock, which is the same node classification task performed in previous works, and predicting the movement of the market index, which is similar to the graph classification task. This is a new way of adapting graph-based learning in stock market prediction. Since market indices consist of individual stocks, we can predict the movement of a market index using a graph classification based approach. The experimental results on both tasks demonstrate the effectiveness of our proposed method. 

The main contributions of this work can be summarized as follows.

\begin{itemize}
	\item We thoroughly investigate the effect of using relations in stock market prediction and find characteristics of more meaningful relation types.
    \item We propose a new method HATS which can selectively aggregate information on different relation types and add the information to each representation. 
	\item We propose graph classification based stock prediction methods. Considering the market index as an entire graph and constituent companies as individual nodes, we predict the movement of a market index using a graph pooling method.
    \item We perform extensive experiments on stocks listed in the S\&P 500 Index. Our experimental results demonstrate that the performance of our method in terms of Sharpe ratio and F1 score was 19.8\% and 3\% higher than the existing baselines, respectively. 
\end{itemize}

The remainder of this paper is organized as follows. In \hyperref[sec:preliminary]{Section 2}, we provide short preliminaries that can be helpful in understanding our work. Detailed descriptions of our proposed framework are provided in \hyperref[sec:method]{Section 3}.  In \hyperref[sec:data]{Section 4}, we explain how we collected the data used in our experiments. We discuss our experimental results in Section \hyperref[sec:exp]{Section 5} and we conclude our work in \hyperref[sec:conclusion]{Section 6}.

%% file: sections/Preliminary.tex
\section{Preliminaries}\label{sec:preliminary}

\paragraph{\textbf{Graph Theory}}
A graph is a powerful data structure which can be used to deal with relational data. Various methods learn meaningful node representations in graphs. In this section, we provide a brief preliminary about a graph based method. Graph $\mathcal{G}$ consists of a set of vertices (nodes) $V$ and edges $E$. If a node is denoted as $v_i \in V$ and $e_{ij} \in E$ is an edge connecting node $i$ and $j$, the Adjacency matrix $A$ is an $n \times n$ matrix with $A_{ij} = w_{ij} > 0$. The degree of a node is the number of edges connected to the node, and is denoted as D where $d_{ii}=\sum_{j}a_{ij}$. Each node can have node features (attributes) X, where $X \in R^{n\times f}$ is a feature matrix. 

The features of nodes change over time in a \textit{spatial-temporal graph} which can be defined using a feature matrix $X \in R^{t\times n \times f}$ where $t$ is the length of time steps. 

\paragraph{\textbf{Graph Neural Networks}}
With a growing interest in utilizing graph-structured data, a large amount of research has been conducted for learning meaningful representations in graphs. Most graph neural networks (GNNs) can be categorized as spectral or non-spectral. 

Spectral graph theory based methods such as GCN \cite{GCNs} utilize convolutional neural networks (CNN \cite{CNNs}) to capture local patterns in graph-structured data. GCN applies a spectral convolution filter to extract information in the Fourier domain.
\begin{equation}
    f_{\theta}(M,x) = UMU^{T}x,
    \tag{2.1} \label{eq:2.1}
\end{equation}
Equation \eqref{eq:2.1} describes a spectral convolution filter $f_{\theta}$ used for graph data $x \in \mathbb{R}^{n}$ (n companies) and a diagonal matrix M. U is the eigenvector matrix of a graph Laplacian matrix.

However, in large graph data, computing eigendecomposition of graph Laplacian is computationally too expensive. To address this problem, \citeauthor{GCNs} approximated spectral filters in terms of Chebyshev polynomials $T_{k}(x)$ up to $k^{th}$ order based on Chebyshev coefficient $\theta_{k}$, which can be defined as follows.
\begin{equation}
    M \approx \sum_{k=0}^{K}\theta_{k}T_{k}(\tilde\wedge), 
    \tag{2.2} \label{eq:2.2}
\end{equation}
where $\tilde\wedge = \frac{2}{\lambda_{max}}\wedge - I$ with $\lambda_{max}$ denotes the largest eigenvalue of graph Laplacian $L$. 

Additionally, Chebyshev coefficients could be represented as $T_{k}(x) = 2xT_{k-1}(x) - T_{k-2}(x)$ with $T_{1}x = x$ and $T_{0}x = 0$. In \cite{GCNs}, GCN is proven to be effective with the parameter setting of K=1. Also, they simply transformed Equation \eqref{eq:2.2} as a fully connected layer with a built-in convolution filter. 

On the other hand, non-spectral approaches directly define convolution operations directly on the graph, utilizing spatially close neighbors. For example, \citeauthor{GraphSAGE} proposed a general framework for sampling and aggregating features from the local neighborhood of a node to generate embeddings. Specifically, features of neighboring nodes are aggregated iteratively using a learnable aggregation function, which is described as follows. 
\begin{equation}
    \tag{2.4} \label{eq:2.4} 
    h^k_{\mathcal{N}(v)} \longleftarrow AGGREGATE(h_u^{k-1}, \forall u \in \mathcal{N}(v)) 
\end{equation}
\begin{equation}
    \tag{2.5} \label{eq:2.5} 
     h^k_v \longleftarrow \sigma \big(W^k \cdot CONCAT(h_u^{k-1},  h^k_{\mathcal{N}(v)}) \big)
\end{equation}
where $h_u^{k}$ denotes the representation of node $u$ at $k$-th iteration and $AGGREGATE$ is a learnable aggregation function. Many proposed methods can be considered as special types of aggregation functions. For example,\citeauthor{GAT} assigned different weights using attention mechanism to aggregate features of neighboring nodes \cite{GAT}. 

Updated node representations can be used in both node classification and graph classification tasks. For a graph classification task, additional layers are needed to sum individual node representations and make graph representations. Graph pooling is a technique used in making graph representations. Numerous works which can effectively aggregate node features have been proposed \cite{SagPool}, \cite{DiffPool}.

GNN methods have proven to achieve state-of-the-art performance in various tasks such as link prediction \cite{linkprediction}, social network community structure prediction \cite{chen2018fastgcn}, and recommendation \cite{recom}.

%% file: sections/Methodology.tex
\section{Methodology\label{sec:method}}
\begin{figure*}[]
    \begin{center}
    \includegraphics[width=16.7cm,height=6.3cm]{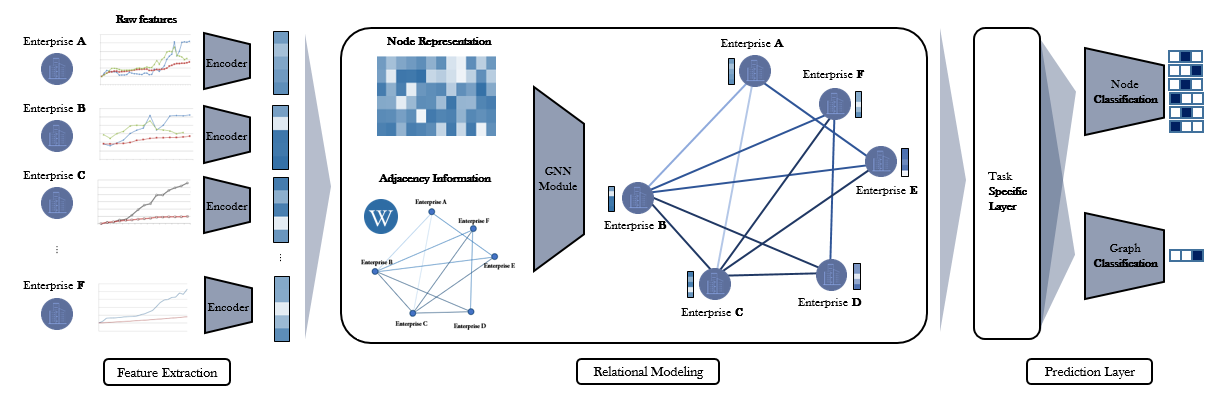}
    \caption{General framework of stock prediction using relational data.}
    \label{fig:gen_framework}
    \end{center}
\end{figure*}

In this section, we will first explain our entire framework. Our framework is based on many different stock market prediction methodologies that use corporate relational data. Knowing how the general framework functions can help in understanding the importance of using relational data in stock prediction. The overall framework is shown in \hyperref[fig:gen_framework]{Fig. 1}. After providing a general description of the framework, we will elaborate on the structure of our method HATS is a new type of relational modeling module.  

\subsection{General framework\label{subsec:method_framework}}

\paragraph{\textbf{Feature Extraction Module}} A stock market graph is a typical type of spatial-temporal graph. If we regard individual stocks (companies) as nodes, each node feature can represent the current state of each company with respect to price movement. Also, node features can evolve over time. As mentioned in \hyperref[sec:intro]{Section 1}, numerous types of data (e.g. historical price, text or fundamental analysis based sources) can be used as an indicator for the movement of a stock price. As data such as raw text or price data are not informative enough, we need a feature extraction module for obtaining meaningful representations of individual companies. In this study, we use only historical price data. 

A feature extraction module is used to represent the current state of a company based on historical movement patterns. In this study, we use LSTM and GRU as our feature extraction modules. LSTM is the most widely used framework in time series modeling and \cite{CIKM-GraphStock} and \cite{Temporal} have also used LSTM as their feature extraction module. For a more detailed description of how node feature vectors are extracted from raw price data, we refer readers to \cite{LSTM-Stock}. We also use GRU as a feature extraction module as it is known to be more efficient than LSTM in time series tasks, and obtains similar performance with appropriate tuning. From our experiments, we found that LSTM performs slightly better than GRU on average. However, it was more difficult to train LSTM especially when a model had more layers. For this reason, we use LSTM for the individual stock prediction task and GRU for the index movement prediction task where an additional graph pooling layer is needed. 

\paragraph{\textbf{Relational Modeling Module} }
A relational modeling module is a node updating function. \citeauthor{MPNNs} considered graph-based learning as information exchange between related nodes \cite{MPNNs}. The main function of graph neural networks is information exchange between neighboring nodes. Information from neighboring nodes is aggregated and then added to each node representation. Information collected from different nodes and relation types needs to be effectively combined. To this end, we propose a new GNN based Hierarchical graph Attention Network for Stock market prediction (HATS) method. Each layer is designed to capture the importance of neighboring nodes and relation types. A detailed description of our proposed method HATS is provided in the below section \hyperref[subsec:HATS]{Subsection}.

\paragraph{\textbf{Task-Specific Module}}
After node representations are updated using relational modeling, the node representations are fed into the task-specific module. Since node representations can be used in various tasks with appropriate modeling, the layer is considered "task-specific." In this study, we performed experiments on the following two graph-based learning tasks: individual stock prediction and market index prediction. Individual stock prediction is similar to the node classification task which was performed in previous researches \cite{CIKM-GraphStock}, \cite{Temporal}. As market indices consist of multiple related stocks, information on the current state of an individual company can be utilized to predict the movement of its index. As recently proposed graph pooling methods can be used to aggregate information of individual nodes to represent an entire graph, they can also be used for the index prediction task. The experimental results in \hyperref[sec:exp]{Section 5} demonstrate that the graph pooling methods outperform all baseline methods. 

In the next subsections, we describe HATS in more detail. We present our method HATS which aggregates information and adds it to node representations. Then, we explain how we use node representations with added information in two different tasks. 

\begin{figure*}[]
    \centering
    \includegraphics[width=16cm,height=9cm]{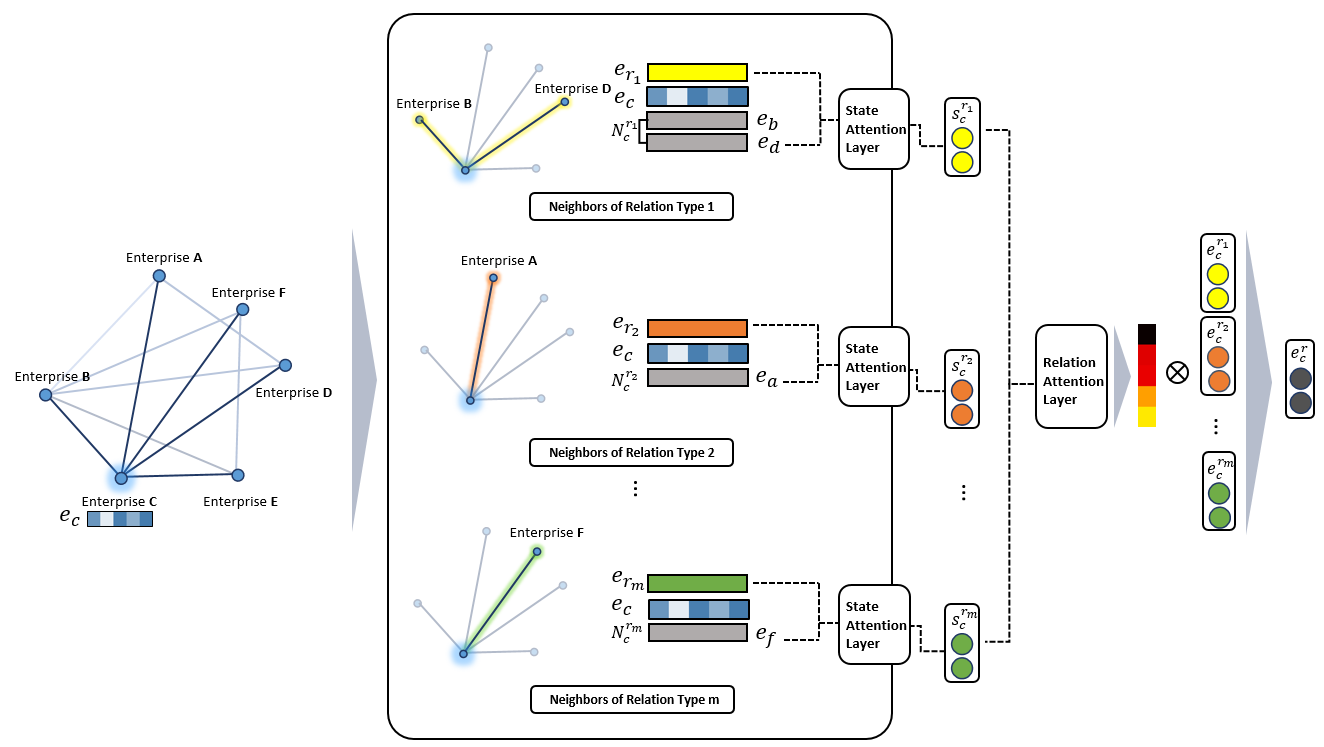}
    \caption{Hierarchical Attention Network for Stock Prediction}
    \label{fig:HATS}
    \centering
\end{figure*}

\subsection{Hierarchical Attention Network\label{subsec:HATS}}

Let us denote a \textit{f}-dimensional feature vector from a feature extraction module of a company \textit{i} at time \textit{t} as $e^t_i \in R^f$. In \hyperref[fig:HATS]{figure 2}, we omit superscript \textit{t} for simplicity, assuming that all the node representation vectors are calculated at time step \textit{t}. We can define edges between different types of relations. For the graph neural network operation, we have to know the set of neighboring nodes for our target node \textit{i} from each relation type. Let us denote the set of neighboring nodes of \textit{i} for relation type m as $N_i^{r_m}$ and the embedding vector of relation type m as $e_{r_m} \in R^d$. Here, \textit{d} is a dimension of a relation type embedding vector. Our goal is to selectively gather information on different relations from neighboring nodes. We want our models to filter information that is not useful for future trend prediction. This process is important because companies have many different types of relationships and some information is not related to movement prediction. 

Attention mechanism is widely used to assign different weight values for information selection. With hierarchically designed attention mechanism, our Hierarchical Attention network for Stock prediction (HATS) selects only meaningful information at each level. Its hierarchical attention network is key in improving performance. The architecture of HATS is shown in \hyperref[fig:HATS]{Fig. 2}.

At the first \textit{state attention layer}, HATS selects important information on the same type of relation from a set of neighboring nodes. The attention mechanism is used to calculate different weights based on the current state (representation) of a neighborhood node. To calculate the state attention scores, we concatenate relation type embedding $e_{r_m}$ and the node representations of \textit{i} and \textit{j} into a vector where $j \in N^{r_m}_i$. If we denote the concatenated vector as $x^{r_m}_{ij}\in R^{(2f+d)}$, the state attention score is calculated as follows: 

\begin{equation} \label{eq:attention}\tag{3.1}
        v_{ij} = x^{r_m}_{ij} W_s + b_s\
\end{equation}
\begin{equation} \label{eq:att_score}\tag{3.2}
    \alpha_{ij}^{r_m} = \frac{\exp (v_{ij})}{\sum_{k} \exp (v_{ik})}\, , 
      \quad \; k\in N^{r_m}_i\, 
\end{equation}
where $W_s \in R^{(2f+d)}$ and $b_s\in R$ are learnable parameters used to calculate the state attention scores. 
With attention weight calculated using \hyperref[eq:attention]{Eq. (3.2)}, we combine all weighted node representations to calculate a vector representation of relation $m$ for company $i$ as \hyperref[eq:att_sum]{Eq. (3.3)}.

\begin{equation} \label{eq:att_sum}\tag{3.3}
    s^{r_m}_{i} = \sum\limits_{j \in N^{r_m}} \alpha_{ij}^{r_m} e_{j}\, 
\end{equation}
With above equation, all the representations of each type of relation are obtained. We selectively gathered information on specific relations from neighboring nodes. A representation can be considered as summarized information of a relation. Vector $s^{r_m}_i$ contains summarized information from relation $m$. For example, the representation of the industry relation summarizes the general state of the industry of our target company. Like human investors, our model should prioritize trading decisions based on summarized information of each relation. The second layer of HATS is designed to continuously assign importance weights to information using another attention mechanism. 

We concatenate the summarized relation information vector $s^{r_m}_i$, representation of the current state of company $e_i$, and the relation type embedding vector $e_{r_m}$ to use $\Tilde{x}^{r_m}_i \in R^{(2f+d)}$ as input for the \textit{relation attention layer}. 

\begin{equation} \label{eq:rel_attention}\tag{3.4}
        \Tilde{v}^{r_m}_{i} = \Tilde{x}^{r_m}_{i} W_r + b_r\
\end{equation}
\begin{equation} \label{eq:rel_att_score}\tag{3.5}
    \Tilde{\alpha}_{i}^{r_m} = \frac{\exp (\Tilde{v}_{i}^{r_m})}{\sum_{k} \exp (\Tilde{v}_{i}^{r_k})}\, , 
      \quad \;|N^{r_k}_i|\ne 0\, 
\end{equation}
$W_r \in R^{(2d+f)}$ and $b_r\in R$ are learnable parameters, and weighted vectors of each relation type are added to form an aggregated relation representation as stated in \hyperref[eq:rel_att_sum]{Eq. (3.6)} which is similar to \hyperref[eq:att_sum]{Eq. (3.3)}.  
\begin{equation} \label{eq:rel_att_sum}\tag{3.6}
    e^r_{i} = \sum_{k} \Tilde{\alpha}_{i}^{(r_k)} s_{i}^{r_k}\, 
\end{equation}
Finally, the representation of a node is added. 

\begin{equation} \label{eq:node_update}\tag{3.7}
    \bar{e}_{i} = e^r_i + e_i\, 
\end{equation}
In the next two subsections, we describe how updated node representations can be used in different tasks.

\subsection{Individual Stock Prediction Layer \label{subsec:ind_pred}}

Like previous works such as \cite{CIKM-GraphStock} and \cite{Temporal}, our model can be applied to the individual stock prediction task. We performed classification on the following three types of labels: \textit{[up, neutral, down]}. A detailed description of the task setting is provided in \hyperref[sec:exp]{Section 5}. For the individual stock prediction task, we added only a simple linear transformation layer. 

\begin{equation} \label{eq:pred}\tag{3.8}
        \hat{Y}_{i} = softmax(\bar{e}_{i} W^n_p + b^n_p)
\end{equation}
where $W^n_p \in R^{d \times l}$, $b^n_p \in R^l$, and l is the number of movement classes.
We trained models on all the corporate relational data using cross-entropy loss. 

\begin{equation} \label{eq:loss}\tag{3.9}
        Loss_{node} = -\sum\limits_{i \in Z_u}\sum\limits_{c=1}^{l}Y_{ic}ln\hat{Y}_{ic}
\end{equation}
where $Y_{ic}$ is a ground truth movement class of company $i$ and $Z_u$ denotes all the companies in our dataset.

\subsection{Graph Pooling for Index Prediction}

A market index consists of multiple stocks chosen based on specific criteria. Let us denote a graph of a specific market index with $n_k$ companies as $\mathcal{G^k}$, where a group of constituent companies of index $k$ is $V^k$ and its updated node representation is $\bar{X} \in R^{n_k \times f}$. To obtain the representation of the entire graph, the features of individual nodes need to be aggregated. Recently, numerous graph pooling methods for aggregation, such as \cite{SagPool} and \cite{DiffPool}, have been proposed. Stock market index data has its own historical price patterns which can be used as features. Therefore, we combine features obtained by graph pooling individual nodes and features directly extracted from historical price data. 

We used mean pooling methods in our experiments to calculate graph representations as follows:

\begin{equation} \label{eq:pooling}\tag{3.9}
        g_p^k = \frac{1}{n_k}\sum\limits_{i \in V^k}\bar{e}_{i}
\end{equation}
where $\bar{e_i}$ is the updated representation of company $i$. By denoting the target index's own feature vector extracted using the feature extraction module as $g^k_e$, the final representation of an entire graph can be obtained by combining the original representation of the graph and the representation obtained by graph pooling as follows.

\begin{equation} \label{eq:g_rep}\tag{3.10}
        g^k = g_p^k + g_e^k
\end{equation}
We also concatenated the two representations; however, this did not have a significant impact on performance. As in the individual stock prediction task, we make predictions using simple linear transformation with $W^g_p \in R^{d \times l}$ and $b^g_p \in R^l$, and train models using cross-entropy loss as follows. 
	
\begin{equation} \label{eq:pred_g}\tag{3.11}
        \hat{Y} = softmax(g^k W^g_p + b^g_p)
\end{equation}
\begin{equation} \label{eq:loss_g}\tag{3.12}
        Loss_{graph} = -\sum\limits_{c=1}^{l}Y_{c}ln\hat{Y}_{c}
\end{equation}
Note that we use the most basic pooling method as this is the first work to apply graph pooling to the stock prediction task. There exists much room for improvement, which we we leave for future work. 

\begin{figure*}\label{fig:task_setting}
    \begin{center}
    \includegraphics[width=17cm,height=4.5cm]{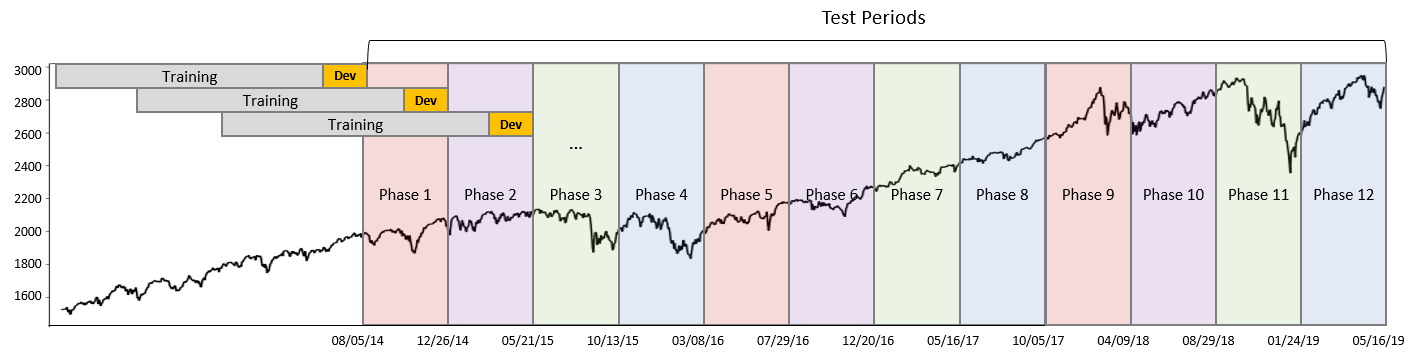}
    \caption{Dataset arrangement for the experiment. A 50-day period is used for evaluation right before the test period. A black line indicates an actual S\&P 500 Index closing price.}
    \end{center}
\end{figure*}

%% file: sections/Data.tex
\section{Data\label{sec:data}}

\subsection{Price-related data}

In this study, we focused on the U.S. stock market, one of the largest markets in the world based on market capitalization. We gathered corporate relational data from a public database which contains information on most of the S\&P 500 companies. Among the S\&P listed companies, there exist some companies without any type of relation with other companies in the database. After removing such companies, the remaining 431 companies were used as our target companies.

We sampled price data for our study from 2013/02/08 to 2019/06/17 (1174 trading days in total). \hyperref[fig:task_setting]{Figure 3.} shows the closing price of the S\&P 500 index, which represents the overall market condition. As shown in \hyperref[fig:task_setting]{Figure 3.}, although the index price has a tendency to go up, there are several crashes in our sample period. We splitted the entire dataset into several phases with varying degrees of volatility to evaluate performance. A more detailed description of the task settings is provided in Section \ref{sec:exp}.

As we described in Section \ref{sec:method}, raw features of price-related data are fed to the feature extraction module. Many different types of raw features such as open price, close price, and volume can be used. In this study, following \cite{LSTM-Stock}, we use historical price change rates as our input. Let $P^t_i$ and $P^{t-1}_i$ be the closing prices of a company \textit{i} at time $t and t-1$, respectively. The price change rate at time \textit{t} is calculated as $R^t_i = \frac{(P^t_i -P^{t-1}_i)}{P^{t-1}_i}$. As our model can predict the movement of a stock price, the price change rate can also be predicted. Therefore, our model can predict the price change rate of the next day $R^{t}_i$, given the sequence of the historical price change rate of a company $[R^{t-l}_i, R^{t-l+1}_i, R^{t-l+2}_i, ... , R^{t-1}_i]$.

\subsection{Corporate Relation data}

The second type of data we used is corporate relational data. Following \citeauthor{Temporal}, we collected corporate relational data from Wikidata \cite{vrandevcic2014wikidata}. Wikidata is a free collaborative knowledge base which contains relations between various types of entities (e.g. person, organization, country). Each entity has an index. If two entities have a relationship, it is considered as {property}. For example, the sentence "Steve jobs founded Apple" is expressed as a triplet \textit{[Apple, Founded by, Steve jobs]}. In terms of graphs, each entity in Wikidata is a node and each property is an edge. Therefore, Wikidata can be understood as a heterogeneous graph with many different types of nodes and edges.

Here, companies are the only node type in which we are interested. However, there are a few types of edges between companies and their connections are very sparse. To address this problem, we utilize meta-path which is commonly used to deal with heterogeneous graphs \cite{metapath2vec}. If Steve Jobs was a board member of Google, we can make the triplet \textit{[Steve jobs, Board Member, Google]}. Combined with the above mentioned relation \textit{[Apple, Founded, Steve jobs]}, the two companies Apple and Google are now connected by the meta-path [Founded by, Board member] and share the node Steve Jobs. In this way, we found that there exist 75 types of relations including direct relations between companies and meta-paths. The entire lists of individual relations and meta-paths used in this study are provided in the appendix. 

One of our main goals is to study the effect of using corporate relational data on stock market prediction performance. There are many ways to define a set of neighboring nodes. We used a meta-path with only 2 hops at maximum to convert an originally heterogeneous graph into a homogeneous graph with only company nodes. Still, methods for building a corporate relational network from a large knowledge base can be much improved, which we leave for future work.

%% file: sections/Experiments.tex
\section{Experiments\label{sec:exp}}

\subsection{Experiment design}
\label{subsec:exp_setting}
\paragraph{\textbf{General settings}} -
As we mentioned in \hyperref[subsec:ind_pred]{Section 3.3}, we divided the training data into the following three classes based on the price change ratio: \textit{[up, neutral, down]}. Specifically, two threshold values were used to divide the training data into the three classes and to assign labels to evaluation and test data. This labeling strategy labels small movements as neutral and uses only the significant movements as directional labels.   

As shown in \hyperref[fig:task_setting]{figure 3}, although the price tends to go up eventually, there exist frequent stock market crashes. To ensure a strategy is profitable, it is important to keep your drawdown at a minimum level. Therefore, we should determine whether models are effective even in a highly volatile period. For this purpose, we divided our entire dataset into 8 smaller datasets that went through different phases, following \cite{bao2017deep}. Each phase consists of 250 days of training, 50 days of evaluation, and 100 days of testing. 

For all the models in our experiments, we used a 50-day lookback period. As we used only the price change ratio as our input feature, the length of the input vector is 50. We used LSTM as a feature extraction module for individual stock prediction, and GRU for the index movement prediction task. We optimized all the models using the Adam optimizer, and tuned the hyper-parameters for each model within a certain range. Specifically, we used a learning rate between 1e-3 and 1e-5, weight decay between 1e-4 and 1e-5, and dropout between 0.1 and 0.9. Relu was used as our activation function. We measured the performance of the models on the evaluation set for each period. We performed early stopping based on F1 score. As the results of the stock prediction task tend to vary widely, all the experiments in this work were repeated five times. The results were averaged to obtain numbers in the table. 

To measure the profitability of the models, we used a trading strategy based on movement prediction. Following \citeauthor{LSTM-Stock}, we made a neutralized portfolio based on the prediction value obtained by models \cite{LSTM-Stock}.Since there are three classes, the prediction vectors from all the models are three dimensional. Values of each dimension represent the predicted probability of each class. We selected 15 companies with the highest up class probability and the long position was taken. For the 15 companies with the highest down class probability, the short position was taken. This method is widely used when creating simple trading strategies for prediction models. We implemented our model in TensorFlow. Our source code and data are available at \href{https://github.com/dmis-lab/hats}{https://github.com/dmis-lab/hats}.

\paragraph{\textbf{Measurement}} 
We evaluated our models based on profitability and classification. In general, creating profitable trading strategies is the ultimate goal of stock movement prediction. Using the trading strategy mentioned above, we used two metrics to calculate profitability. 

\begin{itemize}
\item \textbf{Return} We calculated the return of our portfolio as follows.
\begin{equation*} \label{eq:return}\tag{5.1}
              Return^{t}_{i} = \sum_{i\in F^{t-1}}\frac{(p_i^t- p_i^{t-1})}{p_i^{t-1}} * (-1)^{Action_i^{t-1}}
          \end{equation*}
where $F^{t-1}$ denotes a set of companies included in the portfolio at time $t-1$, and $p_i^t$ denotes the price of stock $i$ at time $t$. $Action_i^{t}$ is a binary value between 0 and 1. $Action_i^{t}$ is 0 if the long position is taken at time $t$ for stock $i$; otherwise, it is 1. 

\item \textbf{Sharpe Ratio} Sharpe ratio is used to measure the performance of an investment compared to its risk. The ratio calculates the earned return in excess of the risk-free rate per unit of volatility (risk) as follows:
\begin{equation*} \label{eq:sharpe}\tag{5.2}
             Sharpe_{a} = \frac{E[R_a - R_f]}{std[R_a-R_f]}
\end{equation*}  
\end{itemize}

where $R_a$ denotes an asset return and $R_f$ is the risk free rate. We used the 13-week Treasury bill to calculate the risk-free rates. 

As price movement prediction is a special type of classification task, we used metrics widely used in classification tasks. 
\begin{itemize}
\item \textbf{Accuracy and F1-Score} These two metrics are the most widely used for measuring classification performance. 

Each prediction can be labeled as True Positive(TP), True Negative(TN), False Positive(FP), or False Negative (FN). Accuracy and F-Score are calculated as follows.
\end{itemize}
\begin{equation*} \label{eq:acc}\tag{5.3}
              Accuracy = \frac{TP+TN}{TP+TN+FP+FN}
          \end{equation*}
\begin{equation*} \label{eq:pr_rc}\tag{5.4}
              Precision = \frac{TP}{TP+FP}, 
              \quad \;Recall = \frac{TP}{TP+FN}
          \end{equation*}
          \begin{equation*} \label{eq:F1}\tag{5.5}
             F1 = 2*\frac{Recall*Precision}{Recall+Precision}
          \end{equation*}

After calculating the F1-score of each class, we averaged all the scores to obtain the macro F1-score.

\paragraph{\textbf{Methods}} - We conducted experiments on the following baseline models. We describe the architecture of each model. We used different combinations of architectures and found that deeper structures generally suffer from overfitting.\\ 
\newline \indent \textit{Baselines without the relational modeling module.}
\begin{itemize}

\item \textbf{MLP} Basic Multi Layer Perceptron model. We used an MLP consisting of 2 hidden layers with 16 and 8 hidden dimensions, respectively, and 1 prediction layer. 

\item \textbf{CNN} We used Convolutional Neural Network as it is known to be fast and as effective as RNN-based models in time series modeling. In our experiments, we used CNN with 4-layers and 2 convolutions and 2 pooling operations. The two convolutional layers with filter sizes of 32 and 8, respectively, and 5 kernels are used for each layer. 

\item \textbf{LSTM} Long Short-Term Memory is one of the most powerful deep learning models for time series forecasting. Many previous works have proven the effectiveness of LSTM. We used a LSTM network with 2 layers  and a hidden size of 128. To train LSTM, we used the RMSProp optimizer which is known to be suitable for RNN-based models. 
\end{itemize}

\textit{Baselines with the relational modeling module.}
\begin{itemize}

\item \textbf{GCN}\cite{CIKM-GraphStock} Basic Graph Convolutional Neural network model. Following \cite{CIKM-GraphStock}, we used a GCN model with two convolution layers and one prediction layer as stated in \hyperref[eq:GCN]{Eq. (5.6)}. All types of relations are used to create an adjacency matrix. 

\item \textbf{GCN-TOP20} We used the same GCN model but we used the edges from only the 20 best performed types of relations in the experiment, described in \hyperref[subsec:method_rel_anal]{subsection 5.2}, to create an adjacency matrix. In other words, we create adjacency matrix with manually selected relations for stock market prediction. By comparing GCN-Top20 with vanilla GCN, we analyzed the effect of using different relations on stock market prediction performance. 

\item \textbf{TGC}\cite{Temporal} Temporal Graph Convolution module for Relational modeling. \citeauthor{Temporal} proposed a general module for stock prediction. This module assigns values to the neighboring nodes of the target company based on the current state of the company and the relations between the nodes and the company. TGC aggregates all the information of a target company from its neighboring nodes while our HATS model summarizes information on different relation types.
\end{itemize}
  
 As mentioned in \hyperref[subsec:method_framework]{Section 3.1}, for all models with a relational modeling module, LSTM is used as a feature extraction module in the individual stock prediction task. In the index movement prediction task, GRU is used as a feature extraction module. The simpler design of GRU makes it easier to train and helps obtain consistent results with deeper model architecture. Therefore, we used GRU as a feature extraction module for all models with the relational modeling module in the index movement prediction task.

\begin{table}[]
\caption{Results of using different relations from Phase 4.}
    \label{table:rel_compare}
\begin{tabular}{cc}
\hline
\multicolumn{2}{c}{\cellcolor[HTML]{C0C0C0}Best 10}                 \\ \hline
\multicolumn{1}{c|}{Relation Type}                         & F1     \\ \hline
\multicolumn{1}{c|}{Industry-Legal form}                   & 0.3276 \\ \hline
\multicolumn{1}{c|}{Industry-Product or material produced} & 0.3251 \\ \hline
\multicolumn{1}{c|}{Parent organization-Owner of}          & 0.325  \\ \hline
\multicolumn{1}{c|}{Owned by-Subsidiary}                   & 0.3247 \\ \hline
\multicolumn{1}{c|}{Parent organization}                   & 0.3247 \\ \hline
\multicolumn{1}{c|}{Founded by-Founded by}                 & 0.3245 \\ \hline
\multicolumn{1}{c|}{Follows}                               & 0.3244 \\ \hline
\multicolumn{1}{c|}{Complies with-Complies with}           & 0.3242 \\ \hline
\multicolumn{1}{c|}{Owner of-Parent organization}          & 0.3241 \\ \hline
\multicolumn{1}{c|}{Subsidiary-Owner of}                   & 0.3241 \\ \hline
\multicolumn{2}{c}{...}                                             \\ \hline
\multicolumn{2}{c}{\cellcolor[HTML]{C0C0C0}Worst 10}                \\ \hline
\multicolumn{1}{c|}{Legal form-Instance of}                & 0.311  \\ \hline
\multicolumn{1}{c|}{Instance of-Legal form}                & 0.3082 \\ \hline
\multicolumn{1}{c|}{Location of formation-Country}         & 0.307  \\ \hline
\multicolumn{1}{c|}{Country-Location of formation}         & 0.3053 \\ \hline
\multicolumn{1}{c|}{Stock Exchange}         & 0.2952 \\ \hline
\multicolumn{1}{c|}{Country of origin-Country}             & 0.2948 \\ \hline
\multicolumn{1}{c|}{Country-Country of origin}             & 0.2886 \\ \hline
\multicolumn{1}{c|}{Country-Country of origin}             & 0.2851 \\ \hline
\multicolumn{1}{c|}{Instance of-Instance of}               & 0.2748 \\ \hline
\multicolumn{1}{c|}{Stock Exchange-Stock Exchange}         & 0.2665 \\ \hline
\end{tabular}
\end{table}

\begin{table*}[]\label{tab:ind_cls}
\caption{Classification accuracy scores on the individual stock prediction task}
\begin{tabular}{cccccccc}
\hline
\multicolumn{8}{|c|}{F1}                                                                                                                                                                                                                                                                                                                                                                                                                                 \\ \hline
\multicolumn{1}{|c|}{}                                & \multicolumn{1}{c|}{MLP}                            & \multicolumn{1}{c|}{CNN}                            & \multicolumn{1}{c|}{LSTM}                       & \multicolumn{1}{c|}{GCN}                            & \multicolumn{1}{c|}{GCN20}                          & \multicolumn{1}{c|}{TGC}                            & \multicolumn{1}{c|}{HATS}                                    \\ \hline
\multicolumn{1}{|c|}{Phase 1}                         & \multicolumn{1}{c|}{0.2876}                         & \multicolumn{1}{c|}{0.3111}                         & \multicolumn{1}{c|}{0.3173}                         & \multicolumn{1}{c|}{0.2874}                         & \multicolumn{1}{c|}{0.3161}                         & \multicolumn{1}{c|}{0.3110}                         & \multicolumn{1}{c|}{\textbf{0.3314}}                         \\ \hline
\multicolumn{1}{|c|}{Phase 2}                         & \multicolumn{1}{c|}{0.2862}                         & \multicolumn{1}{c|}{0.3208}                         & \multicolumn{1}{c|}{0.3228}                         & \multicolumn{1}{c|}{0.3068}                         & \multicolumn{1}{c|}{0.3339}                         & \multicolumn{1}{c|}{0.3088}                         & \multicolumn{1}{c|}{\textbf{0.3347}}                         \\ \hline
\multicolumn{1}{|c|}{Phase 3}                         & \multicolumn{1}{c|}{0.2763}                         & \multicolumn{1}{c|}{0.2938}                         & \multicolumn{1}{c|}{0.3064}                         & \multicolumn{1}{c|}{0.2692}                         & \multicolumn{1}{c|}{\textbf{0.3113}}                & \multicolumn{1}{c|}{0.2237}                         & \multicolumn{1}{c|}{0.3100}                                  \\ \hline
\multicolumn{1}{|c|}{Phase 4}                         & \multicolumn{1}{c|}{0.2810}                         & \multicolumn{1}{c|}{0.3176}                         & \multicolumn{1}{c|}{0.3030}                         & \multicolumn{1}{c|}{0.2940}                         & \multicolumn{1}{c|}{0.3240}                         & \multicolumn{1}{c|}{0.2970}                         & \multicolumn{1}{c|}{\textbf{0.3267}}                         \\ \hline
\multicolumn{1}{|c|}{Phase 5}                         & \multicolumn{1}{c|}{0.2873}                         & \multicolumn{1}{c|}{0.3354}                         & \multicolumn{1}{c|}{0.3333}                         & \multicolumn{1}{c|}{0.3116}                         & \multicolumn{1}{c|}{0.3450}                         & \multicolumn{1}{c|}{0.3329}                         & \multicolumn{1}{c|}{\textbf{0.3496}}                         \\ \hline
\multicolumn{1}{|c|}{Phase 6}                         & \multicolumn{1}{c|}{0.2855}                         & \multicolumn{1}{c|}{0.3265}                         & \multicolumn{1}{c|}{0.3229}                         & \multicolumn{1}{c|}{0.2914}                         & \multicolumn{1}{c|}{0.3140}                         & \multicolumn{1}{c|}{0.2798}                         & \multicolumn{1}{c|}{\textbf{0.3394}}                         \\ \hline
\multicolumn{1}{|c|}{Phase 7}                         & \multicolumn{1}{c|}{0.2876}                         & \multicolumn{1}{c|}{0.3111}                         & \multicolumn{1}{c|}{0.3173}                         & \multicolumn{1}{c|}{0.2874}                         & \multicolumn{1}{c|}{0.3161}                         & \multicolumn{1}{c|}{0.3110}                         & \multicolumn{1}{c|}{\textbf{0.3314}}                         \\ \hline
\multicolumn{1}{|c|}{Phase 8}                         & \multicolumn{1}{c|}{0.2862}                         & \multicolumn{1}{c|}{0.3208}                         & \multicolumn{1}{c|}{0.3228}                         & \multicolumn{1}{c|}{0.3068}                         & \multicolumn{1}{c|}{0.3339}                         & \multicolumn{1}{c|}{0.3088}                         & \multicolumn{1}{c|}{\textbf{0.3347}}                         \\ \hline
\multicolumn{1}{|c|}{Phase 9}                         & \multicolumn{1}{c|}{0.2741}                         & \multicolumn{1}{c|}{0.2390}                         & \multicolumn{1}{c|}{0.2793}                         & \multicolumn{1}{c|}{0.2980}                         & \multicolumn{1}{c|}{0.3160}                         & \multicolumn{1}{c|}{0.2851}                         & \multicolumn{1}{c|}{\textbf{0.3219}}                         \\ \hline
\multicolumn{1}{|c|}{Phase 10}                        & \multicolumn{1}{c|}{0.2529}                         & \multicolumn{1}{c|}{0.2128}                         & \multicolumn{1}{c|}{0.3134}                         & \multicolumn{1}{c|}{0.3002}                         & \multicolumn{1}{c|}{\textbf{0.3272}}                & \multicolumn{1}{c|}{0.2951}                         & \multicolumn{1}{c|}{0.3243}                                  \\ \hline
\multicolumn{1}{|c|}{Phase 11}                        & \multicolumn{1}{c|}{0.2500}                         & \multicolumn{1}{c|}{0.2270}                         & \multicolumn{1}{c|}{0.2997}                         & \multicolumn{1}{c|}{0.2714}                         & \multicolumn{1}{c|}{0.3031}                         & \multicolumn{1}{c|}{0.2577}                         & \multicolumn{1}{c|}{\textbf{0.3091}}                         \\ \hline
\multicolumn{1}{|c|}{Phase 12}                        & \multicolumn{1}{c|}{0.2678}                         & \multicolumn{1}{c|}{0.2921}                         & \multicolumn{1}{c|}{0.2968}                         & \multicolumn{1}{c|}{0.2956}                         & \multicolumn{1}{c|}{0.3299}                         & \multicolumn{1}{c|}{0.3270}                         & \multicolumn{1}{c|}{\textbf{0.3396}}                         \\ \hline
\rowcolor[HTML]{C0C0C0} 
\multicolumn{1}{|c|}{\cellcolor[HTML]{C0C0C0}Average} & \multicolumn{1}{c|}{\cellcolor[HTML]{C0C0C0}0.2769} & \multicolumn{1}{c|}{\cellcolor[HTML]{C0C0C0}0.2923} & \multicolumn{1}{c|}{\cellcolor[HTML]{C0C0C0}0.3113} & \multicolumn{1}{c|}{\cellcolor[HTML]{C0C0C0}0.2933} & \multicolumn{1}{c|}{\cellcolor[HTML]{C0C0C0}0.3225} & \multicolumn{1}{c|}{\cellcolor[HTML]{C0C0C0}0.2948} & \multicolumn{1}{c|}{\cellcolor[HTML]{C0C0C0}\textbf{0.3294}} \\ \hline
\multicolumn{8}{|c|}{Accuracy}                                                                                                                                                                                                                                                                                                                                                                                                                           \\ \hline
\multicolumn{1}{|c|}{Phase 1}                         & \multicolumn{1}{c|}{0.3455}                         & \multicolumn{1}{c|}{0.3540}                         & \multicolumn{1}{c|}{0.3597}                         & \multicolumn{1}{c|}{0.3752}                         & \multicolumn{1}{c|}{0.3700}                         & \multicolumn{1}{c|}{\textbf{0.3811}}                & \multicolumn{1}{c|}{0.3725}                                  \\ \hline
\multicolumn{1}{|c|}{Phase 2}                         & \multicolumn{1}{c|}{0.3342}                         & \multicolumn{1}{c|}{0.3626}                         & \multicolumn{1}{c|}{0.3604}                         & \multicolumn{1}{c|}{0.3735}                         & \multicolumn{1}{c|}{0.3726}                         & \multicolumn{1}{c|}{0.3701}                         & \multicolumn{1}{c|}{\textbf{0.3752}}                         \\ \hline
\multicolumn{1}{|c|}{Phase 3}                         & \multicolumn{1}{c|}{0.3547}                         & \multicolumn{1}{c|}{0.3571}                         & \multicolumn{1}{c|}{0.3771}                         & \multicolumn{1}{c|}{\textbf{0.3860}}                & \multicolumn{1}{c|}{0.3834}                         & \multicolumn{1}{c|}{0.3831}                         & \multicolumn{1}{c|}{0.3859}                                  \\ \hline
\multicolumn{1}{|c|}{Phase 4}                         & \multicolumn{1}{c|}{0.3647}                         & \multicolumn{1}{c|}{0.3855}                         & \multicolumn{1}{c|}{0.3816}                         & \multicolumn{1}{c|}{0.3992}                         & \multicolumn{1}{c|}{0.3897}                         & \multicolumn{1}{c|}{\textbf{0.4059}}                & \multicolumn{1}{c|}{0.3884}                                  \\ \hline
\multicolumn{1}{|c|}{Phase 5}                         & \multicolumn{1}{c|}{0.3208}                         & \multicolumn{1}{c|}{0.3834}                         & \multicolumn{1}{c|}{0.3684}                         & \multicolumn{1}{c|}{0.4191}                         & \multicolumn{1}{c|}{0.4164}                         & \multicolumn{1}{c|}{\textbf{0.4239}}                & \multicolumn{1}{c|}{0.4176}                                  \\ \hline
\multicolumn{1}{|c|}{Phase 6}                         & \multicolumn{1}{c|}{0.3300}                         & \multicolumn{1}{c|}{0.3803}                         & \multicolumn{1}{c|}{0.3841}                         & \multicolumn{1}{c|}{0.3627}                         & \multicolumn{1}{c|}{0.3699}                         & \multicolumn{1}{c|}{0.3716}                         & \multicolumn{1}{c|}{\textbf{0.3869}}                         \\ \hline
\multicolumn{1}{|c|}{Phase 7}                         & \multicolumn{1}{c|}{0.3553}                         & \multicolumn{1}{c|}{0.4309}                         & \multicolumn{1}{c|}{0.4252}                         & \multicolumn{1}{c|}{\textbf{0.4510}}                & \multicolumn{1}{c|}{0.4488}                         & \multicolumn{1}{c|}{0.4477}                         & \multicolumn{1}{c|}{0.4502}                                  \\ \hline
\multicolumn{1}{|c|}{Phase 8}                         & \multicolumn{1}{c|}{0.3537}                         & \multicolumn{1}{c|}{0.3901}                         & \multicolumn{1}{c|}{0.3891}                         & \multicolumn{1}{c|}{0.3993}                         & \multicolumn{1}{c|}{0.4040}                         & \multicolumn{1}{c|}{0.4022}                         & \multicolumn{1}{c|}{\textbf{0.4049}}                         \\ \hline
\multicolumn{1}{|c|}{Phase 9}                         & \multicolumn{1}{c|}{0.3711}                         & \multicolumn{1}{c|}{0.3686}                         & \multicolumn{1}{c|}{0.3511}                         & \multicolumn{1}{c|}{\textbf{0.3955}}                & \multicolumn{1}{c|}{0.3836}                         & \multicolumn{1}{c|}{0.3872}                         & \multicolumn{1}{c|}{0.3923}                                  \\ \hline
\multicolumn{1}{|c|}{Phase 10}                        & \multicolumn{1}{c|}{0.3515}                         & \multicolumn{1}{c|}{0.3484}                         & \multicolumn{1}{c|}{0.3630}                         & \multicolumn{1}{c|}{0.3552}                         & \multicolumn{1}{c|}{0.3623}                         & \multicolumn{1}{c|}{0.3667}                         & \multicolumn{1}{c|}{\textbf{0.3674}}                         \\ \hline
\multicolumn{1}{|c|}{Phase 11}                        & \multicolumn{1}{c|}{0.3813}                         & \multicolumn{1}{c|}{0.3924}                         & \multicolumn{1}{c|}{0.3608}                         & \multicolumn{1}{c|}{0.3823}                         & \multicolumn{1}{c|}{0.3794}                         & \multicolumn{1}{c|}{0.3693}                         & \multicolumn{1}{c|}{\textbf{0.3953}}                         \\ \hline
\multicolumn{1}{|c|}{Phase 12}                        & \multicolumn{1}{c|}{0.3434}                         & \multicolumn{1}{c|}{0.3615}                         & \multicolumn{1}{c|}{0.3584}                         & \multicolumn{1}{c|}{0.3530}                         & \multicolumn{1}{c|}{0.3753}                         & \multicolumn{1}{c|}{\textbf{0.3827}}                & \multicolumn{1}{c|}{0.3802}                                  \\ \hline
\rowcolor[HTML]{C0C0C0} 
\multicolumn{1}{|c|}{\cellcolor[HTML]{C0C0C0}Average} & \multicolumn{1}{c|}{\cellcolor[HTML]{C0C0C0}0.3505} & \multicolumn{1}{c|}{\cellcolor[HTML]{C0C0C0}0.3762} & \multicolumn{1}{c|}{\cellcolor[HTML]{C0C0C0}0.3732} & \multicolumn{1}{c|}{\cellcolor[HTML]{C0C0C0}0.3877} & \multicolumn{1}{c|}{\cellcolor[HTML]{C0C0C0}0.3880} & \multicolumn{1}{c|}{\cellcolor[HTML]{C0C0C0}0.3910} & \multicolumn{1}{c|}{\cellcolor[HTML]{C0C0C0}\textbf{0.3931}} \\ \hline
                                                      &                                                     &                                                     &                                                     &                                                     &                                                     &                                                     &                                                             
\end{tabular}
\end{table*}

\subsection{Analysis of the effect of using relation data}

\label{subsec:method_rel_anal}
We first conducted experiments to investigate the impact of using different types of relations for stock market prediction. The experiments were performed on the individual stock prediction task. To measure the effect of different relations, we used a basic GCN model that cannot distinguish the types of relations. Following \cite{CIKM-GraphStock}, we used a GCN with two convolution layers and one prediction layer, which is defined as follows:  
\begin{equation} 
\label{eq:GCN}\tag{5.6}
	\begin{medsize}
       Y^{GCN} = softmax( \widehat{A}ReLU(\widehat{A}ReLU(\widehat{A}X'W^{(0)})W^{(1)})W^{(2)}) 
	\end{medsize}
\end{equation}
where $\widehat{A}=\Tilde{D}^{-\frac{1}{2}}\Tilde{A}\Tilde{D}^{-\frac{1}{2}} $. Here, $\Tilde{A} = A + I$ is an adjacency matrix with added self-connections, and $\Tilde{D}$ is a degree matrix of $\Tilde{A}$. Therefore, changing the relation type changes the adjacency matrix that is fed into the GCN. We list the 10 best and 10 worst relations and their F1 scores on the test set of Phase 4 in Table 1 \hyperref[table:rel_compare]{Table 1}.

Our key findings are as follows.

\begin{itemize}
	\item \textbf{Using relation data does not always yield good results in stock market prediction.} In our worst cases, using relation data significantly decreased performance. On the other hand, some relation information proved to be helpful in prediction. The best performance is 6\% higher than the worst performance. 
    \item \textbf{Densely connected networks usually have noise.} We confirmed this while analyzing the characteristics of the best and worst relations. Although the number of relations does not affect performance the most, less semantically meaningful relations such as country and stock exchange have very dense networks. Intuitively, densely connected networks carry a considerable amount of noise, which adds irrelevant information to the representations of target nodes. 
\item  \textbf{Manually finding optimal relations is laborious.} Although semantically meaningful relations generally help improve performance,  selecting such relations requires much work and expertise.
\end{itemize}

Based on the above findings, we can conclude that relational information should be selectively chosen when using it for stock market prediction. Furthermore, the framework should be designed to automatically select useful information to minimize the need for manual feature engineering. We conducted experiments on two different tasks to verify the effectiveness of different relational modeling approaches.

\subsection{Individual Stock Prediction}

\renewcommand{\arraystretch}{1.1} 
\begin{table*}\label{tab:ind_prof}
\caption{Profitability results on the individual stock prediction task}
\begin{tabular}{cccccccc}
\hline
\multicolumn{8}{|c|}{Average Daily Return (\%)}                                                                                                                                                                                                                                                                                                                                                                                                           \\ \hline
\multicolumn{1}{|c|}{}                                & \multicolumn{1}{c|}{MLP}                            & \multicolumn{1}{c|}{CNN}                             & \multicolumn{1}{c|}{LSTM}                           & \multicolumn{1}{c|}{GCN}                            & \multicolumn{1}{c|}{GCN20}                          & \multicolumn{1}{c|}{TGC}                            & \multicolumn{1}{c|}{HATS}                                    \\ \hline
\multicolumn{1}{|c|}{Phase 1}                         & \multicolumn{1}{c|}{0.0672}                         & \multicolumn{1}{c|}{-0.0506}                         & \multicolumn{1}{c|}{0.0904}                         & \multicolumn{1}{c|}{-0.0264}                        & \multicolumn{1}{c|}{-0.0103}                        & \multicolumn{1}{c|}{-0.0517}                        & \multicolumn{1}{c|}{\textbf{0.1231}}                         \\ \hline
\multicolumn{1}{|c|}{Phase 2}                         & \multicolumn{1}{c|}{-0.0195}                        & \multicolumn{1}{c|}{0.0929}                          & \multicolumn{1}{c|}{0.1005}                         & \multicolumn{1}{c|}{0.1057}                         & \multicolumn{1}{c|}{0.2435}                         & \multicolumn{1}{c|}{0.1247}                         & \multicolumn{1}{c|}{\textbf{0.1759}}                         \\ \hline
\multicolumn{1}{|c|}{Phase 3}                         & \multicolumn{1}{c|}{0.0029}                         & \multicolumn{1}{c|}{-0.0623}                         & \multicolumn{1}{c|}{0.0454}                         & \multicolumn{1}{c|}{-0.0189}                        & \multicolumn{1}{c|}{0.0246}                         & \multicolumn{1}{c|}{-0.0100}                        & \multicolumn{1}{c|}{\textbf{0.0703}}                         \\ \hline
\multicolumn{1}{|c|}{Phase 4}                         & \multicolumn{1}{c|}{0.0945}                         & \multicolumn{1}{c|}{-0.0578}                         & \multicolumn{1}{c|}{0.1429}                         & \multicolumn{1}{c|}{0.0028}                         & \multicolumn{1}{c|}{0.0385}                         & \multicolumn{1}{c|}{0.0113}                         & \multicolumn{1}{c|}{\textbf{0.1779}}                         \\ \hline
\multicolumn{1}{|c|}{Phase 5}                         & \multicolumn{1}{c|}{-0.0623}                        & \multicolumn{1}{c|}{-0.0673}                         & \multicolumn{1}{c|}{-0.0159}                        & \multicolumn{1}{c|}{-0.0427}                        & \multicolumn{1}{c|}{\textbf{0.0415}}                & \multicolumn{1}{c|}{-0.0202}                        & \multicolumn{1}{c|}{0.0183}                                  \\ \hline
\multicolumn{1}{|c|}{Phase 6}                         & \multicolumn{1}{c|}{-0.0002}                        & \multicolumn{1}{c|}{0.0140}                          & \multicolumn{1}{c|}{0.0400}                         & \multicolumn{1}{c|}{0.0748}                         & \multicolumn{1}{c|}{\textbf{0.0828}}                & \multicolumn{1}{c|}{0.0581}                         & \multicolumn{1}{c|}{0.0726}                                  \\ \hline
\multicolumn{1}{|c|}{Phase 7}                         & \multicolumn{1}{c|}{-0.0081}                        & \multicolumn{1}{c|}{0.0272}                          & \multicolumn{1}{c|}{0.0246}                         & \multicolumn{1}{c|}{0.0201}                         & \multicolumn{1}{c|}{-0.0389}                        & \multicolumn{1}{c|}{-0.0143}                        & \multicolumn{1}{c|}{\textbf{0.0860}}                         \\ \hline
\multicolumn{1}{|c|}{Phase 8}                         & \multicolumn{1}{c|}{0.0319}                         & \multicolumn{1}{c|}{-0.0122}                         & \multicolumn{1}{c|}{0.0742}                         & \multicolumn{1}{c|}{0.0837}                         & \multicolumn{1}{c|}{\textbf{0.2356}}                & \multicolumn{1}{c|}{0.2175}                         & \multicolumn{1}{c|}{0.0662}                                  \\ \hline
\multicolumn{1}{|c|}{Phase 9}                         & \multicolumn{1}{c|}{0.0143}                         & \multicolumn{1}{c|}{-0.0311}                         & \multicolumn{1}{c|}{0.0234}                         & \multicolumn{1}{c|}{0.0375}                         & \multicolumn{1}{c|}{-0.0437}                        & \multicolumn{1}{c|}{-0.0211}                        & \multicolumn{1}{c|}{\textbf{0.0394}}                         \\ \hline
\multicolumn{1}{|c|}{Phase 10}                        & \multicolumn{1}{c|}{0.0040}                         & \multicolumn{1}{c|}{-0.0239}                         & \multicolumn{1}{c|}{-0.0209}                        & \multicolumn{1}{c|}{0.0126}                         & \multicolumn{1}{c|}{0.0164}                         & \multicolumn{1}{c|}{-0.0523}                        & \multicolumn{1}{c|}{\textbf{0.0612}}                         \\ \hline
\multicolumn{1}{|c|}{Phase 11}                        & \multicolumn{1}{c|}{0.0395}                         & \multicolumn{1}{c|}{-0.0106}                         & \multicolumn{1}{c|}{-0.0085}                        & \multicolumn{1}{c|}{0.0597}                         & \multicolumn{1}{c|}{0.0685}                         & \multicolumn{1}{c|}{0.1321}                         & \multicolumn{1}{c|}{\textbf{0.1890}}                         \\ \hline
\multicolumn{1}{|c|}{Phase 12}                        & \multicolumn{1}{c|}{0.0068}                         & \multicolumn{1}{c|}{0.0051}                          & \multicolumn{1}{c|}{0.0222}                         & \multicolumn{1}{c|}{0.0334}                         & \multicolumn{1}{c|}{-0.0241}                        & \multicolumn{1}{c|}{0.0775}                         & \multicolumn{1}{c|}{\textbf{0.0732}}                         \\ \hline
\rowcolor[HTML]{C0C0C0} 
\multicolumn{1}{|c|}{\cellcolor[HTML]{C0C0C0}Average} & \multicolumn{1}{c|}{\cellcolor[HTML]{C0C0C0}0.0142} & \multicolumn{1}{c|}{\cellcolor[HTML]{C0C0C0}-0.0147} & \multicolumn{1}{c|}{\cellcolor[HTML]{C0C0C0}0.0432} & \multicolumn{1}{c|}{\cellcolor[HTML]{C0C0C0}0.0285} & \multicolumn{1}{c|}{\cellcolor[HTML]{C0C0C0}0.0529} & \multicolumn{1}{c|}{\cellcolor[HTML]{C0C0C0}0.0376} & \multicolumn{1}{c|}{\cellcolor[HTML]{C0C0C0}\textbf{0.0961}} \\ \hline
\multicolumn{8}{|c|}{Sharpe Ratio (Annualized)}                                                                                                                                                                                                                                                                                                                                                                                                           \\ \hline
\multicolumn{1}{|c|}{Phase 1}                         & \multicolumn{1}{c|}{2.4410}                         & \multicolumn{1}{c|}{-1.4459}                         & \multicolumn{1}{c|}{2.3553}                         & \multicolumn{1}{c|}{-0.2802}                        & \multicolumn{1}{c|}{-0.1013}                        & \multicolumn{1}{c|}{-0.5029}                        & \multicolumn{1}{c|}{\textbf{2.4796}}                         \\ \hline
\multicolumn{1}{|c|}{Phase 2}                         & \multicolumn{1}{c|}{-1.0063}                        & \multicolumn{1}{c|}{3.2835}                          & \multicolumn{1}{c|}{4.0651}                         & \multicolumn{1}{c|}{2.4700}                         & \multicolumn{1}{c|}{\textbf{4.9007}}                & \multicolumn{1}{c|}{3.0525}                         & \multicolumn{1}{c|}{4.3903}                                  \\ \hline
\multicolumn{1}{|c|}{Phase 3}                         & \multicolumn{1}{c|}{0.1070}                         & \multicolumn{1}{c|}{-1.7872}                         & \multicolumn{1}{c|}{1.0642}                         & \multicolumn{1}{c|}{-0.2477}                        & \multicolumn{1}{c|}{0.2994}                         & \multicolumn{1}{c|}{-0.1796}                        & \multicolumn{1}{c|}{\textbf{1.2503}}                         \\ \hline
\multicolumn{1}{|c|}{Phase 4}                         & \multicolumn{1}{c|}{2.1602}                         & \multicolumn{1}{c|}{-0.6064}                         & \multicolumn{1}{c|}{2.2014}                         & \multicolumn{1}{c|}{0.0289}                         & \multicolumn{1}{c|}{0.3173}                         & \multicolumn{1}{c|}{0.1085}                         & \multicolumn{1}{c|}{\textbf{2.3961}}                         \\ \hline
\multicolumn{1}{|c|}{Phase 5}                         & \multicolumn{1}{c|}{-1.6039}                        & \multicolumn{1}{c|}{-1.7851}                         & \multicolumn{1}{c|}{-0.4455}                        & \multicolumn{1}{c|}{-0.7090}                        & \multicolumn{1}{c|}{\textbf{0.6222}}                & \multicolumn{1}{c|}{-0.4131}                        & \multicolumn{1}{c|}{0.4087}                                  \\ \hline
\multicolumn{1}{|c|}{Phase 6}                         & \multicolumn{1}{c|}{-0.0095}                        & \multicolumn{1}{c|}{0.3565}                          & \multicolumn{1}{c|}{1.1960}                         & \multicolumn{1}{c|}{1.8324}                         & \multicolumn{1}{c|}{2.0390}                         & \multicolumn{1}{c|}{\textbf{2.9435}}                & \multicolumn{1}{c|}{1.6945}                                  \\ \hline
\multicolumn{1}{|c|}{Phase 7}                         & \multicolumn{1}{c|}{-0.4010}                        & \multicolumn{1}{c|}{0.9306}                          & \multicolumn{1}{c|}{0.8354}                         & \multicolumn{1}{c|}{0.4078}                         & \multicolumn{1}{c|}{-0.9107}                        & \multicolumn{1}{c|}{-0.6618}                        & \multicolumn{1}{c|}{\textbf{2.0334}}                         \\ \hline
\multicolumn{1}{|c|}{Phase 8}                         & \multicolumn{1}{c|}{1.0398}                         & \multicolumn{1}{c|}{-0.3917}                         & \multicolumn{1}{c|}{2.1975}                         & \multicolumn{1}{c|}{1.3746}                         & \multicolumn{1}{c|}{3.1870}                         & \multicolumn{1}{c|}{\textbf{4.1305}}                & \multicolumn{1}{c|}{1.4830}                                  \\ \hline
\multicolumn{1}{|c|}{Phase 9}                         & \multicolumn{1}{c|}{0.4915}                         & \multicolumn{1}{c|}{-1.9624}                         & \multicolumn{1}{c|}{0.4905}                         & \multicolumn{1}{c|}{0.7758}                         & \multicolumn{1}{c|}{-0.6896}                        & \multicolumn{1}{c|}{-0.3619}                        & \multicolumn{1}{c|}{\textbf{0.8060}}                         \\ \hline
\multicolumn{1}{|c|}{Phase 10}                        & \multicolumn{1}{c|}{0.6667}                         & \multicolumn{1}{c|}{-1.8774}                         & \multicolumn{1}{c|}{-0.5671}                        & \multicolumn{1}{c|}{0.3263}                         & \multicolumn{1}{c|}{1.3576}                         & \multicolumn{1}{c|}{-1.2023}                        & \multicolumn{1}{c|}{\textbf{1.4382}}                         \\ \hline
\multicolumn{1}{|c|}{Phase 11}                        & \multicolumn{1}{c|}{0.8059}                         & \multicolumn{1}{c|}{-0.7052}                         & \multicolumn{1}{c|}{-0.1983}                        & \multicolumn{1}{c|}{2.5786}                         & \multicolumn{1}{c|}{1.3053}                         & \multicolumn{1}{c|}{3.3379}                         & \multicolumn{1}{c|}{\textbf{3.6146}}                         \\ \hline
\multicolumn{1}{|c|}{Phase 12}                        & \multicolumn{1}{c|}{0.1684}                         & \multicolumn{1}{c|}{0.2055}                          & \multicolumn{1}{c|}{0.6334}                         & \multicolumn{1}{c|}{0.8066}                         & \multicolumn{1}{c|}{-0.8201}                        & \multicolumn{1}{c|}{1.7799}                         & \multicolumn{1}{c|}{\textbf{1.9014}}                         \\ \hline
\rowcolor[HTML]{C0C0C0} 
\multicolumn{1}{|c|}{\cellcolor[HTML]{C0C0C0}Average} & \multicolumn{1}{c|}{\cellcolor[HTML]{C0C0C0}0.4050} & \multicolumn{1}{c|}{\cellcolor[HTML]{C0C0C0}-0.4821} & \multicolumn{1}{c|}{\cellcolor[HTML]{C0C0C0}1.1523} & \multicolumn{1}{c|}{\cellcolor[HTML]{C0C0C0}0.7803} & \multicolumn{1}{c|}{\cellcolor[HTML]{C0C0C0}0.9589} & \multicolumn{1}{c|}{\cellcolor[HTML]{C0C0C0}1.0026} & \multicolumn{1}{c|}{\cellcolor[HTML]{C0C0C0}\textbf{1.9914}} \\ \hline
                                                      &                                                     &                                                      &                                                     &                                                     &                                                     &                                                     &                                                             
\end{tabular}
\end{table*}

\begin{figure*}
    \begin{center}
 \includegraphics[width=14cm,height=5.3cm]{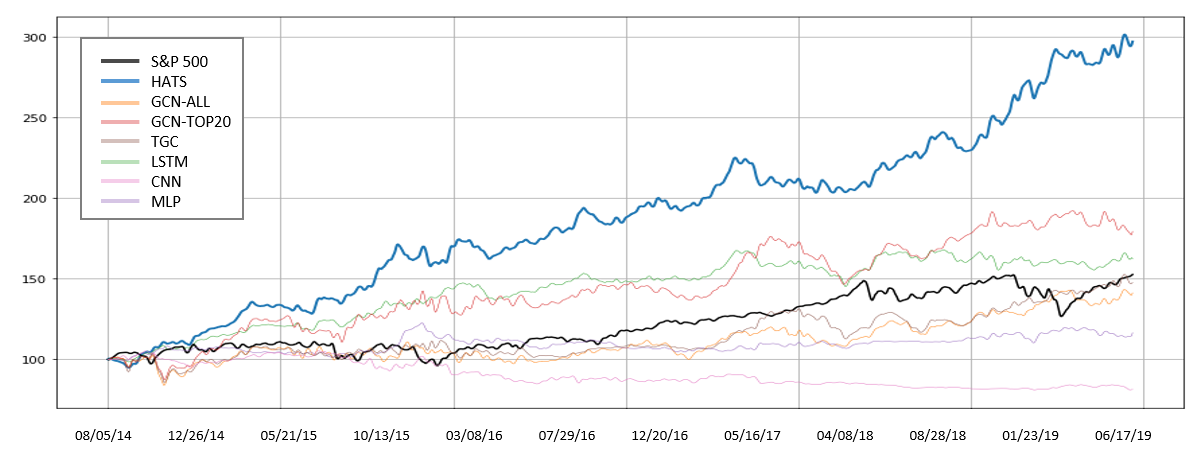}
        \centering
        \label{fig:ind_stock_return}
        \caption{Comparison of different prediction models and their changes in asset value. The asset value is assumed to start at 100.}
    \end{center}    
\end{figure*}

\paragraph{\textbf{Classification Accuracy}} 

The classification accuracy results of the experiments on individual stock market prediction are summarized in \hyperref[tab:ind_cls]{Table 2}. Among the baselines without a relational modeling module, LSTM generally performs the best. Therefore, we compare the results of the models with a relational modeling module and the result of LSTM. In terms of accuracy, all models with a relational modeling module performed better than LSTM. However, not all relational models outperformed LSTM in terms of F1 score. As shown in \hyperref[tab:ind_cls]{Table 2}, only GCN-Top20 and HATS achieved higher F1 scores. What we observed during our experiment is that GCN and TGC tend to make biased predictions on a specific class. By making biased predictions, the GCN and TGC models obtained higher accuracy but lower F1 scores. 

Selectively aggregating information from different relations can help improve F1 scores. Although TGC performed better than vanilla GCN, TGC was outperformed by GCN-Top20 which was trained on manually selected relational data. In contrast, our proposed model HATS generally outperformed all the baselines in terms of both F1 score and accuracy. These results are consistent with the profitability test results which are provided in the following subsection. 

\paragraph{\textbf{Profitability test}} 
The individual stock prediction results on the profitability test are summarized in \hyperref[tab:ind_prof]{Table 3}. We calculated the daily returns of the neutralized portfolio made using the strategy discussed in \hyperref[subsec:exp_setting]{Section 5.1}, and averaged them for each period. On average, GCN-Top20 and HATS obtained the highest average daily return. As mentioned above, GCN-Top20 and HATS outperformed GCN and TGC in terms of F1 score. TGC performed better than vanilla GCN but worse than LSTM. Surprisingly, the Sharpe ratio of GCN-Top20 was lower than that of LSTM and TGC. Without even calculating the Sharpe ratio, we can see in \hyperref[tab:ind_prof]{Table 3} that the expected return results of GCN-Top20 have large variance, which may be attributed to GCN-Top20 using relational data statically. Although relations used for GCN-Top20 are manually selected and expected to improve stock prediction, fixed relations may be useful only in a specific market condition. As GCN cannot assign importance to neighboring nodes based on the market condition and current state of a given node, its results vary widely. By selecting useful information based on the market situation, our HATS model obtains good performance in terms of expected return and Sharpe ratio.

\renewcommand{\arraystretch}{1.1} 
\begin{table*}\label{tab:graph_cls}
\caption{Classification accuracy scores on the market index prediction task}
\begin{tabular}{ccccccc}
\cline{2-7}
\multicolumn{1}{c|}{\textbf{}}     & \multicolumn{1}{c|}{\textbf{S5CONS}} & \multicolumn{1}{c|}{\textbf{S5FINL}} & \multicolumn{1}{c|}{\textbf{S5INFT}} & \multicolumn{1}{c|}{\textbf{S5ENRS}} & \multicolumn{1}{c|}{\textbf{S5UTIL}} & \textbf{Average}                        \\ \hline
\multicolumn{7}{c}{\textbf{F1}}                                                                                                                                                                                                                                                 \\ \hline
\multicolumn{1}{c|}{\textbf{MLP}}  & \multicolumn{1}{c|}{0.2986}          & \multicolumn{1}{c|}{0.3002}          & \multicolumn{1}{c|}{0.2867}          & \multicolumn{1}{c|}{0.2785}          & \multicolumn{1}{c|}{0.2928}          & \cellcolor[HTML]{C0C0C0}0.2913          \\ \hline
\multicolumn{1}{c|}{\textbf{CNN}}  & \multicolumn{1}{c|}{0.3013}          & \multicolumn{1}{c|}{\textbf{0.3157}}          & \multicolumn{1}{c|}{0.3036}          & \multicolumn{1}{c|}{0.3011}          & \multicolumn{1}{c|}{0.3025}          & \cellcolor[HTML]{C0C0C0}0.3049          \\ \hline
\multicolumn{1}{c|}{\textbf{LSTM}} & \multicolumn{1}{c|}{0.3405}          & \multicolumn{1}{c|}{0.2859}          & \multicolumn{1}{c|}{0.3454}          & \multicolumn{1}{c|}{0.3109} & \multicolumn{1}{c|}{0.2942}          & \cellcolor[HTML]{C0C0C0}0.3154          \\ \hline
\multicolumn{1}{c|}{\textbf{GCN}}  & \multicolumn{1}{c|}{0.3410}          & \multicolumn{1}{c|}{0.3040}          & \multicolumn{1}{c|}{0.3423}          & \multicolumn{1}{c|}{0.2848}          & \multicolumn{1}{c|}{0.3111}          & \cellcolor[HTML]{C0C0C0}0.3166          \\ \hline
\multicolumn{1}{c|}{\textbf{TGC}}  & \multicolumn{1}{c|}{0.3322}          & \multicolumn{1}{c|}{0.3051}          & \multicolumn{1}{c|}{0.3391}          & \multicolumn{1}{c|}{0.2736}          & \multicolumn{1}{c|}{0.2911}          & \cellcolor[HTML]{C0C0C0}0.3082          \\ \hline
\multicolumn{1}{c|}{\textbf{HATS}} & \multicolumn{1}{c|}{\textbf{0.3758}} & \multicolumn{1}{c|}{0.3148} & \multicolumn{1}{c|}{\textbf{0.3518}} & \multicolumn{1}{c|}{\textbf{0.3267}}          & \multicolumn{1}{c|}{\textbf{0.3256}} & \cellcolor[HTML]{C0C0C0}\textbf{0.3389} \\ \hline
\multicolumn{7}{c}{\textbf{Accuracy}}                                                                                                                                                                                                                                           \\ \hline
\multicolumn{1}{c|}{\textbf{MLP}}  & \multicolumn{1}{c|}{0.3290}          & \multicolumn{1}{c|}{0.3463}          & \multicolumn{1}{c|}{0.3318}          & \multicolumn{1}{c|}{0.3210}          & \multicolumn{1}{c|}{0.3282}          & \cellcolor[HTML]{C0C0C0}0.3313          \\ \hline
\multicolumn{1}{c|}{\textbf{CNN}}  & \multicolumn{1}{c|}{0.3429}          & \multicolumn{1}{c|}{0.3392}          & \multicolumn{1}{c|}{0.3434}          & \multicolumn{1}{c|}{0.4126}          & \multicolumn{1}{c|}{0.3235} & \cellcolor[HTML]{C0C0C0}0.3537          \\ \hline
\multicolumn{1}{c|}{\textbf{LSTM}} & \multicolumn{1}{c|}{0.3625}          & \multicolumn{1}{c|}{0.3506}          & \multicolumn{1}{c|}{0.3808}          & \multicolumn{1}{c|}{0.4550}          & \multicolumn{1}{c|}{0.3100}          & \cellcolor[HTML]{C0C0C0}0.3718          \\ \hline
\multicolumn{1}{c|}{\textbf{GCN}}  & \multicolumn{1}{c|}{0.3722}          & \multicolumn{1}{c|}{0.3637}          & \multicolumn{1}{c|}{0.3591}          & \multicolumn{1}{c|}{0.4531}          & \multicolumn{1}{c|}{0.3373}          & \cellcolor[HTML]{C0C0C0}0.3771          \\ \hline
\multicolumn{1}{c|}{\textbf{TGC}}  & \multicolumn{1}{c|}{0.4021}          & \multicolumn{1}{c|}{\textbf{0.3699}}          & \multicolumn{1}{c|}{0.3754}          & \multicolumn{1}{c|}{0.4468}          & \multicolumn{1}{c|}{0.3250}          & \cellcolor[HTML]{C0C0C0}0.3819          \\ \hline
\multicolumn{1}{c|}{\textbf{HATS}} & \multicolumn{1}{c|}{\textbf{0.4095}} & \multicolumn{1}{c|}{0.3662} & \multicolumn{1}{c|}{\textbf{0.3834}} & \multicolumn{1}{c|}{\textbf{0.4620}} & \multicolumn{1}{c|}{\textbf{0.3531}}          & \cellcolor[HTML]{C0C0C0}\textbf{0.3948} \\ \hline
\end{tabular}
\end{table*}

\subsection{Market Index Prediction}

As mentioned in \hyperref[sec:data]{section 4}, we gathered price and relational data for 431 companies listed in the S\&P 500. There exist 9 different market indices each representing an industrial sector. We removed four indices with less than 20 constituent companies and have five remaining market indices. The five market indices are as follows: S5CONS (S\&P 500 Consumer Staples Index), S5FINL (S\&P 500 Financials Index), S5INFT (S\&P 500 Information Technology Index), S5ENRS (S\&P 500 Energy Index), S5UTIL (S\&P 500 Utilities Index). As the graph of constituent companies is already sparse, we do not use GCN-Top20 as a baseline. The results are summarized in \hyperref[tab:graph_cls]{table 5}. 

Due to the space constraints, we provide only the averaged results for each index in \hyperref[tab:graph_cls]{table 5}. Furthermore, we did not measure the profitability performance of a neutralized portfolio on the market index prediction task. It is not reasonable to make neutralized portfolio With only five assets as our portfolio selection universe. 

On average, models with a relational modeling module outperformed LSTM on the market index prediction task. However, HATS is the only model that achieved significantly better performance than LSTM in terms of F1 score and accuracy. GCN performed slightly better than LSTM and TGC performed worse than LSTM in terms of F1 score. As we used the same pooling operation for all the models, the differences in performance can be mainly attributed to their relational modeling module. This again proves that HATS is effective in learning node representations for a given task. On the market index prediction task, HATS outperforms all the baselines in terms of F1 score and accuracy on average.

Unexpectedly, the other baselines with the relational modeling module did not perform significantly better than LSTM. The baselines cannot easily select information from different relation types and they use a naive structure to obtain graph representations. Many graph pooling methods such as \cite{SagPool} and \cite{DiffPool} have already been proposed for learning graph representations, and proven to be more effective in many different tasks. We expect that more advanced pooling methods will further improve performance on the market index prediction task. 

\subsection{Case Study}

\begin{figure*}
    \centering
        \includegraphics[width=14cm,height=7cm]{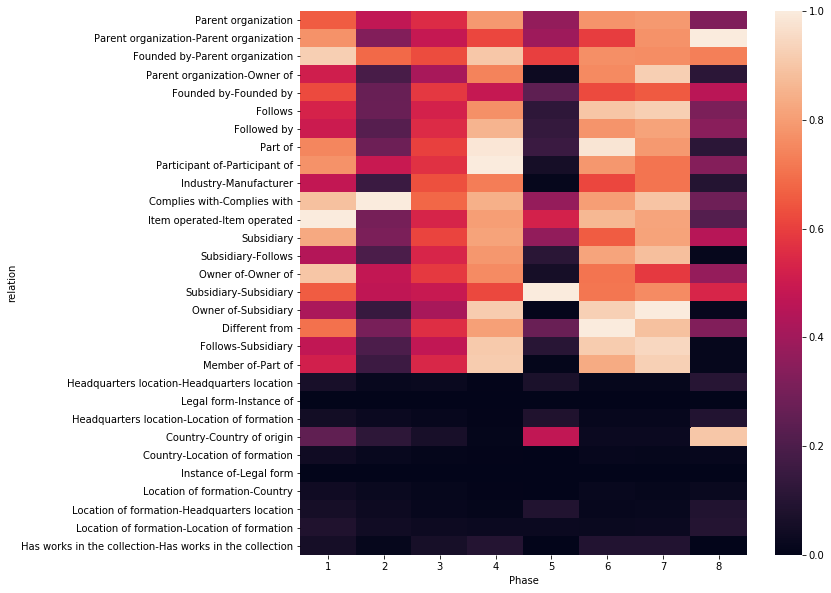}
    
    \caption{Visualization of attention scores of different relations.
    20 relations with highest attention scores on average and 10 relations with the lowest scores on average.}
    \label{fig:rel_attention_score}
\end{figure*}

\paragraph{\textbf{Relation attention scores}} 
In this section, we conduct two case studies to further analyze the decision-making mechanism of HATS. As previously mentioned, HATS is designed to gather information from only useful relations. For our first case study, we calculated the attention score of each relation. By analyzing the relation types with the highest and lowest attention scores, we can understand what types of relations are considered to be important. \hyperref[fig:rel_attention_score]{Fig. 5} shows a visualization of the attention scores of all the relations. We calculated the average attention scores on the test sets from all the phases and selected 20 relations with the highest attention scores and 10 relations with the lowest attention scores. The visualization shown in \hyperref[fig:rel_attention_score]{Fig. 5} is based on the average scores calculated in each test phase. As shown in \hyperref[fig:rel_attention_score]{Fig. 5}, the relations with the highest attention scores are mostly dominant-subordinate relationships such as parent organization-subsidiary relationships. Some relations with the highest scores represent industrial dependencies. On the other hand, most of the relations with the lowest attention scores are geographical features. 

\begin{figure*}[]

    \begin{center}
    \includegraphics[width=17.5cm,height=5cm]{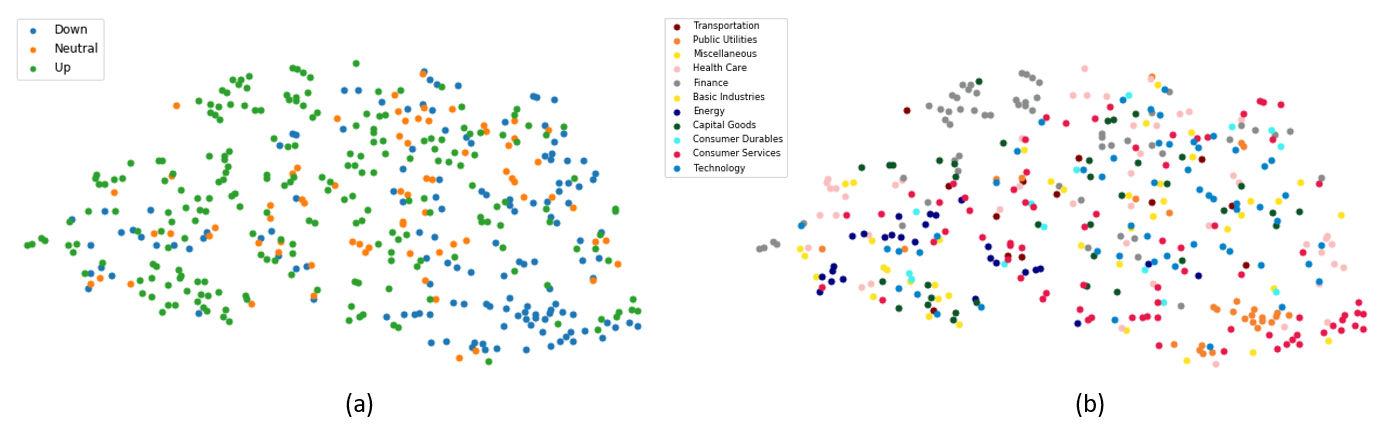}
    \caption{Visualization of node representations using T-SNE.}\label{fig:rep_tsne}
    \end{center}
\end{figure*}

\paragraph{\textbf{Node representation}} 
In studies on graph neural network methods, researchers are interested in representations obtained by GNN. We present the visualization node representation obtained by HATS in \hyperref[fig:rep_tsne]{Fig. 6}. We obtain the representations of all companies on a specific day and use the T-SNE algorithm to map each representation to a two-dimensional space. In Figure 6(a), the movement of a stock on a given day is denoted by any one of the three colors which represent the up/neutral/down labels we used in our experiment. In Figure 6(b),  industries of companies are denoted by different colors. We can find a rough line that separates companies with up labels from companies with down labels in Figure 6(a). It is also interesting that representations of the neutral movement are widely spread. In Figure 6(b), there exists a group of clusters in the same industry. We can find these clusters in any time phase. Although the prices of two stocks in the same industry do not always move in the same direction, the clusters in Figure 6(b) show that HATS learned meaningful representations.

%% file: sections/Conclusion.tex
\section{Conclusion\label{sec:conclusion}}
In this work, we proposed our model HATS which uses relational data in stock market prediction. HATS is designed to selectively aggregate information on different relation types to learn useful node representations. HATS performed the graph related tasks of predicting individual stock prices and predicting market index movement. The experimental results  prove the importance of using proper relational data and show that prediction performance can change dramatically depending on the relation type. The results also show that HATS which automatically selects information to use outperformed all the existing models. 

There exist many possibilities for future research. First, finding a more effective way to construct a corporate network is an important research objective that could be the focus of future studies. In this study, we define the neighborhood of a company as a cluster of companies connected by direct edges or meta-paths with at most 2 hops. However, the way in which we define it could be improved. Furthermore, we used a single database (WikiData) to create a company network. In future work, we could use another source of data and we could even create knowledge graphs from unstructured text of various sources. Applying more advanced pooling methods to obtain graph representations could improve the overall performance of GNN methods on the market index prediction task.

%% file: sections/Appendix.tex

\section{Appendix}

\begin{table*}[]
\caption{List of relation types and their definitions used to make meta-paths. Combinations of relations below are used in our study.}
\begin{tabular}{ccc}
\rowcolor[HTML]{9B9B9B} 
\textbf{Code} & \textbf{Relation Name}                                                                         & \textbf{Description}                                                                                                                                                                                   \\
P17            & Country                                                                                        & sovereign state of this item; don't use on humans                                                                                                                                                      \\ \hline
P31            & Instance of                                                                                    & \begin{tabular}[c]{@{}c@{}}that class of which this subject is a particular example and member \\ (subject typically an individual member with a proper name label)\end{tabular}                       \\ \hline
P112           & Founded by                                                                                     & founder or co-founder of this organization, religion or place                                                                                                                                          \\ \hline
P121           & Item operated                                                                                  & equipment, installation or service operated by the subject                                                                                                                                             \\ \hline
P127           & Owned by                                                                                       & owner of the subject                                                                                                                                                                                   \\ \hline
P131           & \begin{tabular}[c]{@{}c@{}}Located in the \\ administrative \\ territorial entity\end{tabular} & the item is located on the territory of the following administrative entity.                                                                                                                           \\ \hline
P138           & Named after                                                                                    & \begin{tabular}[c]{@{}c@{}}entity or event that inspired the subject's name, \\ or namesake (in at least one language)\end{tabular}                                                                    \\ \hline
P155           & Follows                                                                                        & \begin{tabular}[c]{@{}c@{}}immediately prior item in a series of which the subject is a part \\ {[}if the subject has replaced the preceding item, e.g. political offices, use "replaces"{]}\end{tabular} \\ \hline
P156           & followed by                                                                                    & \begin{tabular}[c]{@{}c@{}}the immediately following item in some series of which the subject is part. \\ Use P1366 if the item is replaced e.g. political offices, states\end{tabular}                \\ \hline
P159           & Headquarters location                                                                          & specific location where an organization's headquarters is or has been situated.                                                                                                                        \\ \hline
P166           & Award received                                                                                 & award or recognition received by a person, organisation or creative work                                                                                                                               \\ \hline
P169           & Chief executive officer                                                                        & highest-ranking corporate officer appointed as the CEO within an organization                                                                                                                          \\ \hline
P176           & Manufacturer                                                                                   & manufacturer or producer of this product                                                                                                                                                               \\ \hline
P355           & Subsidiary                                                                                     & subsidiary of a company or organization, opposite of parent organization                                                                                                                               \\ \hline
P361           & Part of                                                                                        & object of which the subject is a part                                                                                                                                                                  \\ \hline
P400           & Platform                                                                                       & \begin{tabular}[c]{@{}c@{}}platform for which a work was developed or released, \\ or the specific platform version of a software product\end{tabular}                                                 \\ \hline
P414           & Stock Exchange                                                                                 & exchange on which this company is traded                                                                                                                                                               \\ \hline
P452           & Industry                                                                                       & industry of company or organization                                                                                                                                                                    \\ \hline
P463           & Member of                                                                                      & organization or club to which the subject belongs                                                                                                                                                      \\ \hline
P488           & Chairperson                                                                                    & presiding member of an organization, group or body                                                                                                                                                     \\ \hline
P495           & Country of origin                                                                              & country of origin of this item (creative work, food, phrase, product, etc.)                                                                                                                            \\ \hline
P625           & Coordinate location                                                                            & geocoordinates of the subject.                                                                                                                                                                         \\ \hline
P740           & Location of formation                                                                          & location where a group or organization was formed                                                                                                                                                      \\ \hline
P749           & Parent organization                                                                            & parent organization of an organisation, opposite of subsidiaries (P355)                                                                                                                                \\ \hline
P793           & significant event                                                                              & significant or notable events associated with the subject                                                                                                                                              \\ \hline
P1056          & \begin{tabular}[c]{@{}c@{}}Product or \\ material produced\end{tabular}                        & \begin{tabular}[c]{@{}c@{}}material or product produced by a government agency, \\ business, industry, facility, or process\end{tabular}                                                               \\ \hline
P1343          & Described by source                                                                            & dictionary, encyclopaedia, etc. where this item is described                                                                                                                                           \\ \hline
P1344          & Participant of                                                                                 & event a person or an organization was/is a participant in,                                                                                                                                             \\ \hline
P1454          & Legal form                                                                                     & legal form of an organization                                                                                                                                                                          \\ \hline
P1552          & Has quality                                                                                    & the entity has an inherent or distinguishing non-material characteristic                                                                                                                               \\ \hline
P1830          & Owner of                                                                                       & entities owned by the subject                                                                                                                                                                          \\ \hline
P1889          & Different from                                                                                 & item that is different from another item, with which it is often confused                                                                                                                              \\ \hline
P3320          & Board member                                                                                   & member(s) of the board for the organization                                                                                                                                                            \\ \hline
P5009          & Complies with                                                                                  & the product or work complies with a certain norm or passes a test                                                                                                                                      \\ \hline
P6379          & \begin{tabular}[c]{@{}c@{}}Has works \\ in the collection\end{tabular}                         & collection that have works of this artist                                                                                                                                                              \\ \hline
\end{tabular}
\end{table*}

\begin{table*}[]
\caption{List of the relations used in our study. Both direct edges and meta-paths are included in the list.}
\begin{tabular}{cccc}
\hline
\rowcolor[HTML]{9B9B9B} 
\textbf{Relation Index} & \textbf{Relation Combination (Code)} & \textbf{Relation Index} & \textbf{Relation Combination (Code)} \\ \hline
1                       & P1454-P1454                          & 37                      & P452-P176                            \\ \hline
2                       & P159-P159                            & 38                      & P6379-P6379                          \\ \hline
3                       & P1454-P31                            & 39                      & P155-P155                            \\ \hline
4                       & P159-P740                            & 40                      & P495-P17                             \\ \hline
5                       & P17-P495                             & 41                      & P749-P127                            \\ \hline
6                       & P17-P740                             & 42                      & P749-P749                            \\ \hline
7                       & P414-P361                            & 43                      & P4950-P495                           \\ \hline
8                       & P414                                 & 44                      & P495-P740                            \\ \hline
9                       & P452-P452                            & 45                      & P355                                 \\ \hline
10                      & P361-P361                            & 46                      & P355-P155                            \\ \hline
11                      & P127-P749                            & 47                      & P1830-P1830                          \\ \hline
12                      & P1344-P1344                          & 48                      & P112-P749                            \\ \hline
13                      & P127-P127                            & 49                      & P1056-P452                           \\ \hline
14                      & P740-P159                            & 50                      & P1454-P452                           \\ \hline
15                      & P740-P740                            & 51                      & P355-P127                            \\ \hline
16                      & P112-P112                            & 52                      & P176-P452                            \\ \hline
17                      & P155                                 & 53                      & P159-P131                            \\ \hline
18                      & P1056-P1056                          & 54                      & P1830                                \\ \hline
19                      & P1056-P31                            & 55                      & P1830-P127                           \\ \hline
20                      & P127-P355                            & 56                      & P1830-P749                           \\ \hline
21                      & P127-P1830                           & 57                      & P355-P355                            \\ \hline
22                      & P361-P414                            & 58                      & P131-P159                            \\ \hline
23                      & P127                                 & 59                      & P5009-P5009                          \\ \hline
24                      & P156                                 & 60                      & P31-P1056                            \\ \hline
25                      & P31-P1454                            & 61                      & P1830-P355                           \\ \hline
26                      & P452-P31                             & 62                      & P740-P17                             \\ \hline
27                      & P452-P1056                           & 63                      & P740-P495                            \\ \hline
28                      & P463-P463                            & 64                      & P1889                                \\ \hline
29                      & P625-P625                            & 65                      & P749-P112                            \\ \hline
30                      & P361                                 & 66                      & P361-P463                            \\ \hline
31                      & P159-P17                             & 67                      & P155-P355                            \\ \hline
32                      & P17-P159                             & 68                      & P463-P361                            \\ \hline
33                      & P793-P793                            & 69                      & P355-P1830                           \\ \hline
34                      & P166-P166                            & 70                      & P121-P121                            \\ \hline
35                      & P31-P452                             & 71                      & P749-P1830                           \\ \hline
36                      & P452-P1454                           & 72                      & P749                                 \\ \hline
\end{tabular}
\end{table*}